\documentclass[preprint,aps,floatfix,superscriptaddress,linenumbers]{revtex4}

\usepackage{graphicx,epsfig,amssymb,amsmath}
\usepackage{relsize,placeins,float}
\usepackage[dvipsnames]{xcolor}
\usepackage[english]{babel}
\usepackage[percent]{overpic}
\usepackage{ragged2e}
\usepackage{amsthm}
\usepackage{bm}
\usepackage{mathrsfs}
\usepackage{hyperref}

\newtheorem{theorem}{Theorem}[section]

\newtheorem{proposition}[theorem]{Proposition}

\begin{document}

\title{Grünwald--Letnikov Memory Truncation in a Fractional Duffing Oscillator: Coherence Loss and Effective Delay Complexity}

\author{Mattia Coccolo}
\affiliation{Nonlinear Dynamics, Chaos and Complex Systems Group, Departamento de F\'{i}sica,
Universidad Rey Juan Carlos, Tulip\'{a}n s/n, 28933 M\'{o}stoles, Madrid, Spain}

\date{\today}

\begin{abstract}
We investigate the dynamical and analytical consequences of truncating the
Grünwald--Letnikov memory term in a fractional Duffing oscillator. The
truncated memory is treated not merely as a computational approximation, but
as a finite-memory modification of the underlying dynamical system.
We define a coherence-loss time from direct comparisons between the
full-history discrete GL reference and its truncated-memory counterpart, and
use it to extract critical memory horizons in parameter planes involving
the forcing amplitude and the fractional order. The results reveal strongly non-monotonic critical memory horizons,
showing that the retained memory required to preserve coherence depends on the
forcing regime, the fractional order, and the nonlinear sensitivity of the
dynamics.

We also derive a local characteristic equation for the truncated GL kernel and show
that it admits a local delay-type interpretation. In particular, a low-order matching
yields an effective representation in terms of an instantaneous contribution plus a
delayed exponential term, providing a causal local surrogate of the finite-memory
kernel. This local spectral viewpoint motivates a positive-delay exponential
representation of the truncated kernel. The minimum number of positive-delay modes
required to reach a prescribed spectral accuracy defines an operational
delay-complexity measure, \(r_{\min}\). Overall, the truncated GL kernel emerges as an intermediate object
between distributed fractional memory and delay-type dynamics, with a local
spectral structure that is associated with the observed coherence loss and
provides an operational diagnostic of effective delay complexity.
\end{abstract}
\maketitle

\section{Introduction}

Memory effects are a central ingredient in many nonlinear systems arising in physics, engineering, biology, and complex media. In such systems, the present state may depend not only on the instantaneous configuration, but also on the past evolution through nonlocal temporal operators. Fractional derivatives provide a natural framework to describe this type of hereditary behavior, since they introduce distributed memory kernels extending over the history of the system \cite{Oldham1974,Podlubny1999,Kilbas2006}. Among the different formulations, the Gr\"unwald--Letnikov (GL) derivative is especially useful from the numerical point of view, because it leads directly to a discrete convolution over past states \cite{Podlubny1999,Kilbas2006}.

 For clarity, the term \emph{fractional Duffing oscillator} is used throughout
this work in its standard sense: a Duffing oscillator containing a
non-integer-order fractional derivative, here represented through a
Gr\"unwald--Letnikov memory term. The fractional order therefore describes a
temporal nonlocality and an associated memory effect in the damping term.

This terminology should be distinguished from that of a \emph{fractal Duffing
oscillator}. In that context, the term ``fractal'' may refer to formulations
involving fractal derivatives, fractal media or spaces, or other explicitly
fractal geometric or scale-dependent structures
\cite{Feng2021,ElDib2024}. Such ingredients are not part of the present
formulation. The model studied here is a fractional Duffing oscillator with a
standard Euclidean state space and a Gr\"unwald--Letnikov temporal memory
operator; no fractal derivative, fractal geometry, or fractal-space formulation
is introduced.

In nonlinear oscillators, fractional damping and fractional memory can produce dynamical effects that differ substantially from those associated with classical local dissipation. The interplay between nonlinearity, forcing, and memory may modify stability, resonance, transient dynamics, and the onset of complex behavior. Duffing-type systems provide a canonical setting in which these effects can be investigated, and previous works have shown that fractional damping can strongly reshape both asymptotic and transient responses \cite{Ruzziconi2011,Coccolo2024,CoccoloTransient}. At the same time, the numerical implementation of fractional memory is costly: the GL derivative requires the storage and evaluation of a history-dependent sum whose length increases with time.

This difficulty naturally motivates the use of truncated fractional memory. In
the GL formulation, truncation replaces the full memory sum by a finite-horizon
convolution, so that only the most recent portion of the past contributes to
the dynamics. This idea is related to the short-memory principle
\cite{Podlubny1999}. However, in nonlinear forced systems, truncation should
not be regarded only as a computational shortcut. By cutting the memory tail,
one modifies the kernel itself and therefore changes the effective dynamical
system. This raises a central question: when does the truncated-memory dynamics
remain coherent with the full-memory dynamics, and when does the finite memory
horizon become dynamically insufficient?

The short-memory principle and its numerical consequences have been analyzed
in different fractional integration schemes. Deng~\cite{Deng}, for instance,
studied the short-memory principle in combination with an Adams-type
predictor--corrector approach and provided an error analysis for fractional
differential equations. More recent variants, including improved or piecewise
memory principles, aim to reduce the computational burden while controlling
the approximation error~\cite{Ma,Gong}. In particular, improved short-memory
GL strategies have been used to study bifurcation structures in fractional
Duffing systems~\cite{Ma,Wu2019}. {Related finite-memory and fixed-memory-length versions of the
Gr\"unwald--Letnikov operator have also been considered in the literature as a
way of reducing the computational cost of long-memory fractional simulations;
see, for example, Ref.~\cite{Hamri:2016}. More generally, alternative nonlocal operators with different
memory kernels have also been proposed in the fractional-calculus literature,
including formulations with exponential memory~\cite{FuWu2021}. However, the present work differs from these approaches in two ways. First,
we do not merely assess the numerical accuracy of memory truncation or propose
an alternative fractional operator with a different memory kernel. Instead, we
study the dynamical consequences of truncating the standard GL memory in a
nonlinear forced Duffing oscillator. Second, we treat the truncated GL
operator as a modified dynamical object and quantify its coherence loss and
effective delay complexity.

Fast convolution algorithms, sum-of-exponentials approximations, and
kernel-compression techniques provide a complementary route to reducing the
cost of fractional-memory simulations by approximating or accelerating the
history convolution \cite{Schadle2006,Jiang2017,Baffet2017}. Their primary objective is usually computational
efficiency while preserving the target fractional operator. In contrast, the
present work treats finite GL truncation itself as a modified finite-memory
operator and investigates the associated dynamical consequences.
}

The novelty of this viewpoint can be summarized in three main points.
First, the truncated GL convolution is not used only as a computational
device to accelerate fractional simulations. Instead, it is interpreted as a
finite-memory dynamical operator, and we study when its trajectories remain
coherent with, or depart from, the full-memory GL reference. Second, we
introduce the coherence-loss time \(T_{\rm loss}\) and the critical memory
horizon \(T_m^{\rm crit}\) as operational diagnostics to quantify the dynamical
validity of memory truncation in a nonlinear forced oscillator. These
quantities are not standard local discretization errors, but trajectory-level
measures of the time over which a finite-memory realization preserves the
full-memory dynamics. Third, we analyze the local spectral structure of the
truncated GL kernel and connect it with delay-type representations. In
particular, the positive-delay complexity \(r_{\min}\) measures how many
causal exponential modes are required to reproduce the finite-memory kernel
within a prescribed spectral tolerance. This perspective differs from
conventional fractional Duffing studies, which mainly focus on the effects of
fractional damping on resonance, stability, or chaotic transitions, and from
standard GL approximation studies, whose main objective is usually the numerical
approximation of a continuous fractional derivative.

{ We address this problem for a periodically forced Duffing oscillator with GL fractional memory. Throughout the paper, three distinct objects must be kept separate: the underlying continuous fractional Duffing equation, the full-history discrete GL approximation used as the numerical reference over the finite simulation window, and its truncated finite-memory counterpart. The truncated realization is compared directly with the full-history discrete GL reference by monitoring trajectory errors, coherence-loss times,
critical memory horizons, and parameter-dependent transitions. The memory horizon is written as $T_m=Mh$, where $M$ is the number of retained GL coefficients and $(h)$ is the time step. This quantity plays a dual role: it controls the computational cost of the approximation, but also defines the temporal support of the modified memory operator.
}

A second objective of this paper is to clarify the analytical meaning of the truncated GL kernel. After linearization, the truncated memory term leads to a characteristic equation containing a finite weighted sum of exponentials. This structure is reminiscent of delay equations, which are a classical framework for systems with temporal nonlocality \cite{Bellman1963,Hale1993}. In Duffing-type systems, time-delayed feedback and
delayed interactions are known to modify resonance, predictability, and transition
mechanisms \cite{Cantisan2020,Coccolo2021Delay}. However, this delay-type structure should not be confused with a single physical time delay. {Within the low-order local spectral and moment-matching framework considered
here, a pure one-exponential reduction may be insufficient to capture the
relevant local spectral structure of the truncated kernel. Instead, the
truncated GL kernel is interpreted as a finite distributed-memory operator
that may admit, locally, effective delay-type surrogates in terms of
instantaneous and delayed exponential contributions.
}

This observation motivates a further analysis: we study whether the truncated GL kernel can be approximated by a finite superposition of positive delayed exponentials. The resulting minimum number of required positive-delay modes provides an operational measure of the local ``delay-complexity'' of the truncated
fractional memory at the prescribed spectral tolerance. Numerical scans in the parameter space spanned by the fractional order $q$ and
the retained memory horizon $T_m$ show that this complexity depends
nontrivially on both quantities. In particular, kernels with large fractional order are often highly compressible, whereas intermediate orders and large memory horizons may require a much richer delay-like representation.

The main message of this work is therefore twofold. First, truncating fractional memory is not merely a numerical simplification: it produces a structurally modified dynamical system whose agreement with the full-memory model depends on the forcing regime, the fractional order, and the retained memory horizon. Second, the truncated GL kernel forms an intermediate object between full fractional memory and delay-type dynamics. It retains a distributed-memory structure over a finite temporal support, while its local spectral form can be interpreted through a finite collection of effective delayed exponential contributions.

The paper is organized as follows. In Sec.~II we introduce the fractional
Duffing oscillator and its truncated GL formulation. In Sec.~III we derive the
local characteristic equation associated with the truncated memory kernel and
discuss its delay-type interpretation. Section~IV presents direct numerical
comparisons between full and truncated memory dynamics, including the
coherence-loss time and an amplitude--phase diagnostic. Section~V analyzes
coherence-loss maps and critical memory thresholds in the $(f,T_m)$ plane,
while Sec.~VI presents complementary scans in the $(q,T_m)$ plane.
Section~VII discusses the local spectral interpretation obtained from the
characteristic equation. Section~VIII analyzes the positive-delay exponential
representation of the truncated kernel and the associated delay-complexity
maps. Section~IX discusses the scope and limitations of the proposed
framework and clarifies its relation to existing approximation approaches.
Finally, Sec.~X summarizes the main conclusions.

\section{Fractional and truncated model}

We consider a periodically forced Duffing oscillator with fractional damping. At the
continuous level, the reference model can be written as
\begin{equation}
\ddot{x}(t)+\alpha x(t)+\beta x^3(t)+c\,{}_{0}D_t^q x(t)
=
f\cos(\omega t),
\label{eq:duffing_full}
\end{equation}
where \(0<q<1\), \(\alpha\) and \(\beta\) are the linear and nonlinear stiffness
coefficients, \(c\) measures the strength of the fractional damping term, and \(f\)
and \(\omega\) are the forcing amplitude and frequency, respectively. Here
\({}_{0}D_t^q\) denotes a continuous fractional derivative, understood in the
Riemann--Liouville sense, with lower terminal \(t=0\)
\cite{Podlubny1999,Kilbas2006}. Under the usual assumptions on the initial data, or
after the corresponding initialization correction, the same numerical construction
may also be related to the Caputo formulation.

{
In the present work it is important to distinguish clearly between three levels:
(i) the continuous fractional Duffing equation above;
(ii) its full-history discrete Gr\"unwald--Letnikov realization over the finite
simulation window, which serves as the numerical reference throughout the paper;
and (iii) the truncated finite-memory GL realization, in which only the most
recent \(M+1\) coefficients are retained. The dynamical comparisons reported
below are always made between items (ii) and (iii), not between the truncated
model and the exact continuous fractional solution.
}

The focus of the present work, however, is not the introduction of a new continuous
fractional derivative. Rather, we study the dynamical consequences of using the
Gr\"unwald--Letnikov discrete realization of the fractional operator and, in
particular, of truncating its memory. For a uniform time step \(h>0\), \(t_n=nh\),
the full-history discrete GL approximation of the fractional derivative is
\begin{equation}
{}_{0}D_t^q x(t_n)
\approx
D_h^{q,\mathrm{GL}}x_n
=
\frac{1}{h^q}
\sum_{k=0}^{n}
(-1)^k
\binom{q}{k}
x_{n-k},
\label{eq:GL_full_discrete}
\end{equation}
where \(x_n=x(t_n)\) and
\begin{equation}
\binom{q}{k}
=
\frac{\Gamma(q+1)}
{\Gamma(k+1)\Gamma(q-k+1)}.
\label{eq:gen_binomial}
\end{equation}
This expression is a discrete convolution over the available past states and
provides a natural numerical formulation for the study of memory truncation
\cite{Podlubny1999,Kilbas2006}. Thus, in the present paper the symbol
\(D_h^{q,\mathrm{GL}}\) denotes the {\it full-history discrete}
Gr\"unwald--Letnikov approximation used in the numerical implementation,
whereas \({}_{0}D_t^q\) denotes the underlying continuous fractional derivative.

The finite-memory version is obtained by retaining only the most recent \(M+1\)
terms of the GL convolution:
\begin{equation}
D_h^{q,\mathrm{GL},M}x_n
=
\frac{1}{h^q}
\sum_{k=0}^{\min(n,M)}
(-1)^k
\binom{q}{k}
x_{n-k}.
\label{eq:GL_truncated_discrete}
\end{equation}
The corresponding memory horizon is
\begin{equation}
T_m=Mh.
\label{eq:memory_horizon}
\end{equation}
Operationally, \(T_m\) represents the time span of the past trajectory that is
retained by the finite-memory model once \(n>M\). Contributions from times
earlier than approximately \(t_n-T_m\) are then discarded. Thus, decreasing
\(T_m\) does not only reduce the computational cost, but also shortens the
temporal range over which the system ``remembers'' its previous states.

The truncated model studied in this work is therefore the finite-memory GL
realization of the fractional Duffing oscillator, evaluated on the numerical grid:
\begin{equation}
\ddot{x}(t_n)+\alpha x_n+\beta x_n^3
+
c\,D_h^{q,\mathrm{GL},M}x_n
=
f\cos(\omega t_n),
\label{eq:duffing_truncated}
\end{equation}
Equation~\eqref{eq:duffing_truncated} should not be interpreted as a new
continuous fractional Duffing equation. It is the finite-memory GL realization of
Eq.~\eqref{eq:duffing_full}. In this sense, the truncation modifies the discrete
memory kernel and defines the effective finite-memory dynamics analyzed
throughout the paper.

{
The numerical benchmark used for comparison is the corresponding
{\it full-history discrete GL reference over the finite simulation window},
obtained by replacing \(D_h^{q,\mathrm{GL},M}x_n\) in
Eq.~\eqref{eq:duffing_truncated} with \(D_h^{q,\mathrm{GL}}x_n\), i.e., by
retaining all previously available grid values at each time step within the
simulated interval, or equivalently by taking \(M\geq n\) at each time step.
}

In all numerical simulations reported below we fix \(\alpha=-1\), \(\beta=1\), and
\(c=0.3\). Thus, the finite-memory GL realization actually used in the truncated
simulations is
\begin{equation}
\ddot{x}(t_n)-x_n+x_n^3+0.3\,D_h^{q,\mathrm{GL},M}x_n
=
f\cos(\omega t_n),
\label{eq:duffing_fixed_parameters}
\end{equation}
while the full-history discrete GL reference is obtained by replacing
\(D_h^{q,\mathrm{GL},M}x_n\) with \(D_h^{q,\mathrm{GL}}x_n\).

The corresponding conservative Duffing potential is
\begin{equation}
V(x)=\frac{\alpha}{2}x^2+\frac{\beta}{4}x^4
=
-\frac{1}{2}x^2+\frac{1}{4}x^4,
\label{eq:duffing_potential}
\end{equation}
which is the standard symmetric double-well potential, with minima at
\(x=\pm 1\) and a barrier at \(x=0\). The fractional GL term, its truncated
version, and the effective delay representations modify the memory-dependent
nonconservative contribution, but they do not change this underlying
conservative potential.

{
For a finite simulation ending at time \(t_N=Nh\), the full-history discrete GL
reference corresponds to retaining all grid values available from the simulated
past at each time step, i.e., using the sum \(k=0,\dots,n\) at time \(t_n\).
This is the reference object against which the truncated trajectories are
compared throughout the paper. It should be distinguished from the exact
continuous infinite-memory fractional solution associated with
Eq.~\eqref{eq:duffing_full}. The latter is recovered only in the appropriate
continuous limit, namely by combining the full-history discrete GL construction
with a consistent refinement \(h\to 0\) under the standard assumptions for GL
convergence. By contrast, the present numerical ``full-memory'' benchmark is
always a finite-window, full-history {\it discrete} GL reference.
}

\subsection{Discrete truncation error and convergence considerations}

The truncation error considered in this work is defined at the level of the
discrete GL convolution. For a fixed time step \(h\), the difference between
the full-history discrete GL approximation and its finite-memory counterpart at
time \(t_n\) is
\begin{equation}
\delta_M(t_n)
=
D_h^{q,\mathrm{GL}}x_n
-
D_h^{q,\mathrm{GL},M}x_n
=
\frac{1}{h^q}
\sum_{k=M+1}^{n}
w_k x_{n-k},
\qquad n>M,
\label{eq:discrete_tail_error}
\end{equation}
where \(w_k=(-1)^k\binom{q}{k}\). If \(n\leq M\), the two operators coincide
exactly and \(\delta_M(t_n)=0\). Thus, for fixed \(h\), the truncated operator
coincides with the full-history discrete GL operator once \(M\geq n\) at a
given time step, and over a finite observation interval \(0\leq t\leq t_N\)
the full-history discrete approximation is recovered when \(M\geq N\).

For bounded trajectories, \(|x_n|\leq X\), the discarded tail can be bounded as
\begin{equation}
|\delta_M(t_n)|
\leq
\frac{X}{h^q}
\sum_{k=M+1}^{n}|w_k|.
\label{eq:tail_bound_general}
\end{equation}
For \(0<q<1\), the GL coefficients have an algebraically decaying tail,
\begin{equation}
|w_k|\sim C_q k^{-q-1},
\qquad k\to\infty,
\label{eq:weights_asymptotic}
\end{equation}
where \(C_q\) is a positive \(q\)-dependent constant. Consequently, the
discarded tail satisfies the estimate
\begin{equation}
|\delta_M(t_n)|
\lesssim
C_q X h^{-q}M^{-q}
=
C_q X T_m^{-q},
\label{eq:tail_bound_Tm}
\end{equation}
up to a \(q\)-dependent prefactor. This estimate should be understood as a
bound on the instantaneous operator error, not as a direct bound on the
trajectory error in a nonlinear system. In the Duffing oscillator, the
discarded tail acts as a deterministic perturbation that may be amplified by
the nonlinear dynamics, which motivates the coherence-loss diagnostics
introduced below.

The estimate in Eq.~\eqref{eq:tail_bound_Tm} provides a rigorous algebraic
control of the perturbation introduced by memory truncation at the operator
level. More precisely, for bounded trajectories, the difference between the
full-history and truncated GL memory terms decreases as \(T_m^{-q}\), up to a
(q)-dependent prefactor. This result concerns the direct perturbation of the
memory operator itself. It does not, by itself, yield a uniform bound on the
difference between the corresponding nonlinear trajectories, because the latter
also depends on the finite-time sensitivity of the underlying dynamics.

The continuous and discrete limits should therefore be distinguished. Under
the standard regularity assumptions required for the Gr\"unwald--Letnikov
approximation, the full-history discrete operator \(D_h^{q,\mathrm{GL}}\)
converges to the continuous fractional derivative \({}_{0}D_t^q\) as
\(h\to0\), provided that the number of retained past values increases
consistently with the observation time. By contrast, if the physical memory
horizon \(T_m=Mh\) is kept finite while \(h\to0\), the limiting object is not
the full-memory fractional derivative but a finite-memory version of the
fractional operator. This is precisely the regime analyzed in the present
work: memory truncation is treated as a finite-memory modification of the
discrete GL dynamics, and its validity is assessed dynamically through
\(T_{\rm loss}\), \(T_m^{\rm crit}\), and the associated coherence-loss maps.

{
For brevity, after this point we occasionally use the shorter expression
``full-memory GL reference'', but it should always be understood in the precise
sense of a {\it full-history discrete GL reference over the finite simulation
window}, not as the exact continuous fractional solution.
}

Regarding the numerical scheme, the simulations use a fixed time step \(h\)
and the standard recursive computation of the GL weights. The finite-memory
simulations are compared against the corresponding full-history discrete GL
reference computed with the same time step and initial data. The stability
issue addressed in this work is consequently a dynamical one: we ask whether
the finite-memory trajectory remains close to the full-history discrete
reference over the observation window.

A general stability theorem for the nonlinear forced fractional Duffing
equation is beyond the scope of the present study. Instead, numerical
reliability is assessed by monitoring the phase-space error \(E(t)\), the
coherence-loss time \(T_{\rm loss}\), and the critical memory horizon
\(T_m^{\rm crit}\). In this framework, an early loss of coherence indicates
that the discarded memory tail has become dynamically relevant, even if the
instantaneous truncation error is small.

To implement the GL convolution, we compute the weights recursively. Writing
\begin{equation}
w_k = (-1)^k\binom{q}{k},
\label{eq:weights_definition}
\end{equation}
the coefficients satisfy
\begin{equation}
w_0=1,
\qquad
w_k = w_{k-1}\left(1-\frac{q+1}{k}\right),
\qquad k\geq 1,
\label{eq:weights_recursive}
\end{equation}
which is the form used in the numerical codes. This recursive structure is
especially convenient in parameter sweeps, since it avoids repeated evaluations
of the gamma function.

The central problem addressed in this work is the following: given the
full-history discrete GL dynamics generated by \(D_h^{q,\mathrm{GL}}\), how
large must the memory horizon \(T_m\) be in order for the truncated
finite-memory model generated by \(D_h^{q,\mathrm{GL},M}\) to remain
dynamically coherent with it? As will be shown below, the answer depends
strongly on the forcing regime and on the fractional order. In particular, the
truncation error is not only a matter of local approximation accuracy, but
also of dynamical sensitivity, since different parameter regions amplify the
effect of the neglected memory tail in different ways.

\subsection{Choice of the truncation length}

The truncation length \(M\), or equivalently the memory horizon \(T_m=Mh\),
is not selected from a universal a priori rule. Its appropriate value depends
on the required accuracy, the observation time, and the dynamical regime of
the nonlinear oscillator. In this work, \(M\) is therefore treated as a control
parameter. We scan different values of \(T_m\) and determine whether the
finite-memory trajectory remains coherent with the full-history discrete GL
reference over the prescribed observation interval.

The practical criterion used below is based on the coherence-loss time
\(T_{\rm loss}\). For a prescribed tolerance \(\varepsilon_{\rm loss}\) and a
target time \(T_{\rm target}\), we define the critical memory horizon as
\begin{equation}
T_m^{\rm crit}
=
\min\left\{T_m:\,T_{\rm loss}(T_m)\geq T_{\rm target}\right\},
\label{eq:Tmcrit_general}
\end{equation}
with the dependence on other parameters, such as \(f\) or \(q\), made explicit
in the corresponding parameter scans. Thus, \(T_m^{\rm crit}\) is the smallest
retained memory horizon for which the truncated dynamics remains coherent with
the full-history discrete GL reference up to the required target time.

Increasing \(M\) improves the approximation of the discrete GL memory by
reducing the discarded tail, but it also increases the computational cost,
since each time step requires a longer convolution sum and more past states
must be stored. Conversely, decreasing \(M\) reduces the computational burden
but may remove a dynamically relevant part of the memory kernel. In nonlinear
regimes, this discarded tail can be amplified by the dynamics, producing an
early loss of coherence. Therefore, the choice of \(M\) affects not only the
instantaneous numerical accuracy of the memory term, but also the dynamical
reliability of the finite-memory trajectory over long times.

In the following section, we analyze the truncated operator from a local
analytical viewpoint. After linearization, the finite GL sum leads to a
characteristic equation involving a finite weighted sum of exponential terms.
This structure provides a natural bridge with delay-type characteristic
equations. However, the truncated GL operator should not be interpreted as a
single physical delay. Instead, we show that the finite-memory kernel admits
local spectral approximations in terms of effective exponential contributions,
and we later quantify how many positive-delay exponentials are required to
represent its local spectral action.

{ \subsection{Numerical implementation details}

All time-domain simulations are performed using a fixed-step Euler--Cromer
scheme. Introducing \(v=\dot{x}\), the acceleration at time \(t_n\) is evaluated
as
\begin{equation}
a_n=
-\alpha x_n-\beta x_n^3
-c\,D_h^{q,\mathrm{GL},M}x_n
+f\cos(\omega t_n),
\label{eq:euler_cromer_acceleration}
\end{equation}
for the truncated model, with \(D_h^{q,\mathrm{GL},M}\) replaced by
\(D_h^{q,\mathrm{GL}}\) for the full-history discrete GL reference. The velocity
and position are then updated according to
\begin{equation}
v_{n+1}=v_n+h\,a_n,
\qquad
x_{n+1}=x_n+h\,v_{n+1}.
\label{eq:euler_cromer_update}
\end{equation}
Thus, the GL memory term is evaluated explicitly at the current time level from
the states available up to \(t_n\), while the position update uses the newly
computed velocity.

Unless otherwise stated, all trajectories are initialized with
\begin{equation}
x(0)=0,
\qquad
v(0)=0.
\label{eq:initial_conditions}
\end{equation}
No external history function is prescribed for \(t<0\). Instead, the discrete GL
convolution starts at \(t=0\) and includes only the states available within the
simulation window. Consequently, at the initial steps the convolution is
evaluated over the available partial history, as indicated by the upper limit
\(\min(n,M)\) in Eq.~\eqref{eq:GL_truncated_discrete}.

Tables~\ref{tab:numerical_settings_time} and~\ref{tab:numerical_settings_spectral}  summarize the numerical settings used in the
different analyses. In the coherence-loss maps, the critical memory horizon is
extracted as the smallest sampled value of \(T_m=Mh\) satisfying
\(T_{\rm loss}\geq T_{\rm target}\). The curves superimposed on the maps in
Figs.~3 and 4 are not theoretical fits: they are moving-average guides to the
eye obtained from the numerically extracted threshold values. A window of seven
points is used in Fig.~3, whereas a window of five points is used in Fig.~4,
after excluding unresolved threshold points.

For the local spectral analysis in Fig.~\ref{fig:local_spectral_maxima}, the late-time maxima are extracted
from the discrete trajectory after discarding the first \(70\%\) of the
integration interval. A point \(x_k\) is classified as a local maximum when
\(x_{k-1}<x_k\) and \(x_k\geq x_{k+1}\). The spectral residual is evaluated over
a uniformly sampled frequency grid.

For the positive-delay kernel fitting, the complex spectral domain is sampled on
a rectangular \((\sigma,\omega)\) grid. The fitting amplitudes \(A_0\) and
\(A_j\) are unconstrained, whereas positivity of the delays is enforced through
\(\tau_j=\exp(\theta_j)\). For each number of exponential modes, several random
initializations are used and the fit with the smallest relative error is
retained. Points for which the prescribed tolerance is not reached with
\(r\leq r_{\max}\) are recorded as unresolved.

\begin{table*}[t]
\centering
\small

\caption{Time-domain numerical settings.}
\label{tab:numerical_settings_time}
\renewcommand{\arraystretch}{1.18}
\setlength{\tabcolsep}{5.5pt}
\begin{tabular}{|p{0.18\textwidth}|p{0.33\textwidth}|p{0.36\textwidth}|}
\hline
\textbf{Analysis} &
\textbf{Parameters and grid} &
\textbf{Numerical and post-processing settings} \\
\hline
{\textbf{Common settings}} &
\(\alpha=-1\), \(\beta=1\), \(c=0.3\); \(h=0.01\);
\(x(0)=0\), \(v(0)=0\) &
Fixed-step Euler--Cromer scheme. The GL term is evaluated explicitly from
the states available at time \(t_n\). Full-history and truncated runs use the
same time step and initial data. \\
\hline
Fig.~1 &
\(q=0.70\), \(f=0.30\), \(\omega=1.20\);
\(M=1000,12000\); \(t_{\rm final}=150\) &
Coherence threshold:
\(\varepsilon_{\rm loss}=5\times10^{-2}\). \\
\hline
Fig.~3 &
\(q=0.70\), \(\omega=1.20\);
\(f\in[0.10,0.50]\), 120 points;
\(M\in[10,15000]\), 120 points &
\(T_{\rm target}=142.5\). Seven-point moving-average guide to the finite
threshold values. \\
\hline
Fig.~4 &
\(f=0.30\), \(\omega=1.20\);
\(q\in[0.50,0.95]\), 120 points;
\(M\in[10,15000]\), 120 points &
\(T_{\rm target}=142.5\). Five-point moving-average guide to the finite
threshold values. \\
\hline
Fig.~5 &
\(h=0.01\), \(t_{\rm final}=150\), and
\(T_{\rm target}=142.5\). The same parameter slices as in
Figs.~3 and 4 are used. &
Threshold-sensitivity checks for \(T_m^{\rm crit}(f)\) and
\(T_m^{\rm crit}(q)\);
\(\varepsilon_{\rm loss}=0.025,\,0.05,\,0.10\) \\
\hline
\end{tabular}
\end{table*}

\begin{table*}[t]
\centering
\small

\caption{Spectral and positive-delay fitting settings.}
\label{tab:numerical_settings_spectral}
\renewcommand{\arraystretch}{1.18}
\setlength{\tabcolsep}{5.5pt}
\begin{tabular}{|p{0.18\textwidth}|p{0.33\textwidth}|p{0.36\textwidth}|}
\hline
\textbf{Analysis} &
\textbf{Parameters and grid} &
\textbf{Numerical and post-processing settings} \\
\hline
 Fig.~6 &
 $q=0.90$, $M=12000$, $T_m=120$;
 $f\in[0.05,0.55]$, $750$ points;
 $t_{\rm final}=300$ &
 The first $70\%$ of each trajectory is discarded. A local maximum satisfies
 $x_{k-1}<x_k$ and $(x_k\geq x_{k+1})$. The residual is sampled on
 $(\omega\in[0,2.5])$ using $5000$ points. \\
 \hline
 Fig.~7 &
 $h=0.01$;
 $q\in[0.10,0.95]$, $M\in[500,15000]$;
$100\times100$ parameter grid &
 Spectral grid:
 $\sigma\in[0,0.08]$, $\omega\in[0,0.80]$, $24\times24$ points.
 Twelve random starts per $r$ with logarithmically sampled initial delays; $\varepsilon_{\rm fit}=10^{-2}$;
 $r_{\max}=9$. The resulting map is interpreted as an operational and potentially conservative
estimate near difficult fitting regions. Unresolved cases are those for which no
\(r\leq9\) reaches the prescribed tolerance. \\
 \hline
\end{tabular}
\end{table*}

}

\section{Local analytical approximation of the truncated memory}

In this section we derive a local analytical approximation for the truncated Grünwald--Letnikov (GL) memory term introduced in Sec.~II. The goal is not to replace the original truncated dynamics by an exactly equivalent delay equation, but rather to identify the local spectral structure induced by the finite-memory kernel and to interpret it in terms of effective exponential contributions. This will later provide a useful bridge between the truncated fractional dynamics and the transition mechanisms observed numerically.

\subsection{Linearized equation and characteristic form}

We begin from the truncated Duffing equation

\begin{equation}
\ddot{x}(t)+\alpha x(t)+\beta x^3(t)+c\,D_h^{q,\mathrm{GL},M}x(t)=f\cos(\omega t),
\label{eq:duffing_trunc_local}
\end{equation}
with
\begin{equation}
D_h^{q,\mathrm{GL},M}x(t)
=
\frac{1}{h^q}\sum_{k=0}^{M}(-1)^k\binom{q}{k}x(t-kh).
\label{eq:GL_trunc_local}
\end{equation}

To study the local spectral effect of the truncated memory, we consider the linearization around a reference state $x^\ast$. In the Duffing case, the local restoring coefficient is
\begin{equation}
b(x^\ast)=\alpha+3\beta(x^\ast)^2,
\label{eq:local_b}
\end{equation}
and for the parameter values used in the simulations, this local stiffness becomes
\[
b(x^*)=-1+3(x^*)^2.
\]
The corresponding linearized equation is

\begin{equation}
\ddot{x}(t)+b\,x(t)+c\,D_h^{q,\mathrm{GL},M}x(t)=0.
\label{eq:linearized_trunc}
\end{equation}
Seeking modal solutions of the form
\begin{equation}
x(t)=e^{\lambda t},
\label{eq:modal_ansatz_local}
\end{equation}
one obtains
\begin{equation}
x(t-kh)=e^{\lambda t}e^{-\lambda kh},
\end{equation}
and therefore

\begin{equation}
D_h^{q,\mathrm{GL},M}e^{\lambda t}
=
e^{\lambda t}\,\frac{1}{h^q}\sum_{k=0}^{M}(-1)^k\binom{q}{k}e^{-\lambda kh}.
\label{eq:GL_on_exp}
\end{equation}
Substituting into Eq.~\eqref{eq:linearized_trunc}, we obtain the exact characteristic equation
\begin{equation}
\lambda^2+b+c\,h^{-q}\sum_{k=0}^{M}(-1)^k\binom{q}{k}e^{-\lambda kh}=0.
\label{eq:exact_characteristic}
\end{equation}

This expression already reveals an important point. The truncated GL operator does not produce a polynomial characteristic equation in the usual sense; instead, it yields a finite weighted sum of exponential factors. In this respect, the truncated memory kernel has a structure that is naturally related to delay-type characteristic equations. However, the weights and time scales are inherited from the fractional GL discretization, so this structure should not be interpreted a priori as a standard delay equation with a single physical delay.

\subsection{Small-\texorpdfstring{$|\lambda h|$}{|lambda h|} expansion}

To obtain a local approximation, we consider the regime
\begin{equation}
|\lambda h|\ll 1,
\label{eq:small_lambda_h}
\end{equation}
and expand the exponential factors in Eq.~\eqref{eq:exact_characteristic}. Let
\begin{equation}
w_k=(-1)^k\binom{q}{k},
\label{eq:wk_local}
\end{equation}
and define the discrete moments
\begin{equation}
S_n=\sum_{k=0}^{M}w_k\,k^n
=
\sum_{k=0}^{M}(-1)^k\binom{q}{k}k^n.
\label{eq:Sn_local}
\end{equation}
Using
\begin{equation}
e^{-\lambda kh}
=
1-\lambda kh+\frac{(\lambda kh)^2}{2}+\mathcal{O}\!\left((\lambda h)^3\right),
\label{eq:exp_series_local}
\end{equation}
we obtain
\begin{equation}
\sum_{k=0}^{M}w_k e^{-\lambda kh}
\approx
S_0-\lambda h\,S_1+\frac{(\lambda h)^2}{2}S_2+\mathcal{O}\!\left((\lambda h)^3\right).
\label{eq:kernel_expansion_local}
\end{equation}
Substituting this into Eq.~\eqref{eq:exact_characteristic} yields
\begin{equation}
\lambda^2+b+c\,h^{-q}
\left(
S_0-\lambda h\,S_1+\frac{(\lambda h)^2}{2}S_2
\right)
=0,
\label{eq:char_local_poly}
\end{equation}
or, grouping powers of $\lambda$,
\begin{equation}
\left[b+c\,h^{-q}S_0\right]
-\left[c\,h^{1-q}S_1\right]\lambda
+\left[1+\frac{c}{2}h^{2-q}S_2\right]\lambda^2
\approx 0.
\label{eq:char_grouped_local}
\end{equation}

Thus, at this level, the truncated memory induces local corrections to the effective stiffness, damping-like term, and inertial coefficient. However, while Eq.~\eqref{eq:char_grouped_local} is useful as a polynomial approximation, it does not retain the exponential structure that is naturally present in Eq.~\eqref{eq:exact_characteristic}. For this reason, it is useful to examine reduced exponential representations of the truncated kernel itself, while keeping in mind that a representation involving too few exponential
terms may fail to capture the local spectral structure of the truncated kernel
in a dynamically meaningful way.

\subsection{Minimal exponential approximation of the kernel}

We seek a local approximation of the truncated kernel in the form
\begin{equation}
\sum_{k=0}^{M}w_k e^{-\lambda kh}
\approx
A_1 + A_2 e^{-\lambda \tau_d},
\label{eq:kernel_delay_ansatz}
\end{equation}
where $A_1$, $A_2$, and $\tau_d$ are effective coefficients to be determined. This is the simplest two-term exponential representation capable of reproducing the kernel expansion up to second order in $\lambda$.

Expanding the right-hand side of Eq.~\eqref{eq:kernel_delay_ansatz}, we get
\begin{equation}
A_1 + A_2 e^{-\lambda \tau_d}
=
(A_1+A_2)
-\lambda (A_2\tau_d)
+\frac{\lambda^2}{2}(A_2\tau_d^2)
+\mathcal{O}(\lambda^3).
\label{eq:delay_ansatz_expanded}
\end{equation}
Matching Eqs.~\eqref{eq:kernel_expansion_local} and \eqref{eq:delay_ansatz_expanded} gives
\begin{align}
A_1+A_2 &= S_0,
\label{eq:match_A1A2}
\\
A_2\tau_d &= hS_1,
\label{eq:match_tau}
\\
A_2\tau_d^2 &= h^2S_2.
\label{eq:match_tau2}
\end{align}
From Eqs.~\eqref{eq:match_tau} and \eqref{eq:match_tau2} one obtains
\begin{equation}
\tau_d = h\,\frac{S_2}{S_1},
\label{eq:tau_d_general}
\end{equation}
while
\begin{equation}
A_2=\frac{S_1^2}{S_2},
\qquad
A_1=S_0-\frac{S_1^2}{S_2}.
\label{eq:A1A2_general}
\end{equation}

The discrete moments can be simplified explicitly. Using standard identities for
generalized binomial coefficients, one finds
\begin{align}
S_0 &= (-1)^M\binom{q-1}{M},
\label{eq:S0_local}
\\
S_1 &= q\,(-1)^M\binom{q-2}{M-1}.
\label{eq:S1_local}
\end{align}
To compute \(S_2\), we use the identity
\begin{equation}
k^2=k(k-1)+k.
\label{eq:k2_split}
\end{equation}
Therefore,
\begin{equation}
S_2
=
\sum_{k=0}^{M}w_k\,k^2
=
\sum_{k=0}^{M}w_k\,k(k-1)+\sum_{k=0}^{M}w_k\,k.
\label{eq:S2_split_local}
\end{equation}
The first sum is the second factorial moment of the truncated GL weights, so
\begin{equation}
S_2
=
q(q-1)(-1)^M\binom{q-3}{M-2}+S_1,
\qquad M\geq 2.
\label{eq:S2_local}
\end{equation}

Hence,
\begin{align}
\frac{S_1}{S_0} &= \frac{qM}{q-1},
\label{eq:S1S0_local}
\\
\frac{S_2}{S_1}
&=
1+\frac{q-2}{(q-1)(M-1)}
=
1+\frac{2-q}{(1-q)(M-1)}.
\label{eq:S2S1_local}
\end{align}
It follows that
\begin{equation}
\tau_d
=
h\,\frac{S_2}{S_1}
=
h\left[
1+\frac{q-2}{(q-1)(M-1)}
\right],
\label{eq:tau_d_closed}
\end{equation}
which is positive for \(0<q<1\) and \(M\geq 2\).

Using Eqs.~\eqref{eq:match_tau}--\eqref{eq:match_tau2}, the corresponding
amplitudes are
\begin{equation}
A_2=\frac{hS_1}{\tau_d}=\frac{S_1^2}{S_2},
\qquad
A_1=S_0-\frac{S_1^2}{S_2}.
\label{eq:A1A2_general_corrected}
\end{equation}

Therefore, for \(|\lambda h|\ll 1\), the finite GL kernel admits a minimal
local spectral representation as the sum of an instantaneous contribution and
a single effective exponential term. In contrast with the pure one-delay
approximation discussed below, the corresponding parameter \(\tau_d\) is
positive for \(0<q<1\). However, this representation remains local and
low-order: it should not be interpreted as a global reduction of the truncated
GL kernel to a single physical delay.

\begin{proposition}
For \(|\lambda h|\ll 1\), the truncated Grünwald--Letnikov kernel admits the
local approximation
\begin{equation}
\sum_{k=0}^{M}(-1)^k\binom{q}{k}e^{-\lambda kh}
=
A_1+A_2e^{-\lambda\tau_d}
+\mathcal{O}\!\left((\lambda h)^3\right),
\label{eq:proposition_kernel}
\end{equation}
where \(\tau_d\), \(A_1\), and \(A_2\) are given by
Eqs.~\eqref{eq:tau_d_closed} and \eqref{eq:A1A2_general_corrected},
with \(S_0\), \(S_1\), and \(S_2\) defined in
Eqs.~\eqref{eq:S0_local}--\eqref{eq:S2_local}.
\end{proposition}

\subsection{Equivalent local characteristic equation}

Replacing the finite GL kernel in Eq.~\eqref{eq:exact_characteristic} by its local approximation \eqref{eq:proposition_kernel}, we arrive at
\begin{equation}
\lambda^2+b+c\,h^{-q}\left(A_1+A_2e^{-\lambda\tau_d}\right)=0.
\label{eq:char_after_kernel}
\end{equation}
Introducing
\begin{align}
b_{\mathrm{eff}} &= b+c\,h^{-q}A_1,
\label{eq:beff_local}
\\
K_{\mathrm{eff}} &= c\,h^{-q}A_2,
\label{eq:Keff_local}
\end{align}
the characteristic equation can be written compactly as
\begin{equation}
\lambda^2+b_{\mathrm{eff}}+K_{\mathrm{eff}}e^{-\lambda\tau_d}=0.
\label{eq:equivalent_local_characteristic}
\end{equation}
{
Equation~\eqref{eq:equivalent_local_characteristic} provides a minimal local
exponential representation of the truncated GL memory, formally analogous to
a delay-type characteristic equation. The approximation is not intended as a
global dynamical equivalence, but as a local tool to interpret the spectral
role of the truncated kernel and to provide a diagnostic associated with
transition regions.
}
\subsection{Interpretation}

Several remarks are in order.

First, Eq.~\eqref{eq:equivalent_local_characteristic} shows that the truncated
GL operator is not merely a finite-memory version of the full fractional
derivative in a numerical sense. At the linearized spectral level, it behaves
as an effective combination of an instantaneous contribution and a single
exponential contribution. In the present second-order local matching, the
corresponding effective delay \(\tau_d\) is positive for \(0<q<1\). This means
that, once an instantaneous term is included, the truncated GL kernel admits a
causal local surrogate of the form
\(A_1+A_2e^{-\lambda\tau_d}\).

However, this result should be interpreted with care. The parameter \(\tau_d\)
is not a physical delay built into the original Duffing oscillator, but an
effective quantity obtained by matching the low-order expansion of the
truncated GL kernel in the Laplace domain. Therefore,
Eq.~\eqref{eq:equivalent_local_characteristic} is not a global reduction of
the finite-memory fractional system to a single delayed differential equation.
Rather, it is a local spectral approximation valid in the regime
\(|\lambda h|\ll 1\).
{
This also clarifies the relation with the pure one-delay approximation. Within
the specific low-order local spectral expansion and moment-matching framework
used here, a single exponential contribution without an instantaneous term
cannot reproduce the relevant low-order moments of the truncated GL kernel
with a positive effective delay; the formal matching instead yields a negative
value of the associated delay parameter. This result should not be interpreted
as ruling out every possible one-delay approximation in a broader sense, nor
does it imply non-causality of the original truncated GL scheme, which is
constructed exclusively from present and past samples. By allowing the kernel
to be split into an instantaneous part plus a delayed contribution, the
minimal two-term ansatz considered here yields a positive effective delay.
Thus, the relevant conclusion is not that the truncated GL kernel is
equivalent to a single physical delay, but that its present low-order local
moment structure requires more flexibility than a pure one-delay exponential
description.
}

Second, this provides a natural conceptual bridge between three different
types of temporal nonlocality:
\begin{enumerate}
\item the full fractional case, characterized by a distributed memory extending
over the whole past;
\item the truncated fractional case, characterized by a distributed memory with
finite support;
\item a local delay-type representation, in which the finite-memory kernel is
approximated by effective instantaneous and delayed contributions.
\end{enumerate}

Third, the usefulness of Eq.~\eqref{eq:equivalent_local_characteristic} lies
in the fact that it gives a first local delay-type interpretation of the
truncated kernel, but it does not exhaust its spectral structure. The finite
GL memory still contains a distributed set of weighted past contributions, and
a single delayed exponential---even when supplemented by an instantaneous
term---may be insufficient to reproduce that structure accurately over the
selected spectral region of interest. This motivates the positive-delay
multi-exponential representations introduced below. In that setting, the
central question is no longer whether one formal effective delay can be
defined, but how many positive-delay exponential modes are required to
reproduce the truncated GL kernel within a prescribed spectral tolerance. This
question leads directly to the definition of the minimum delay complexity
\(r_{\min}\).

\section{Numerical comparison between full and truncated memory}

{
We now compare the truncated-memory GL dynamics directly with the corresponding
full-history discrete GL reference over the same finite simulation window. The
purpose of this comparison is not merely to estimate a numerical error, but to
show that memory truncation may produce a genuine dynamical departure when the
retained memory horizon is not sufficiently large.
}

For this purpose, we consider the phase-space error
\begin{equation}
E(t)=
\sqrt{
\left[x_{\rm full}(t)-x_M(t)\right]^2+
\left[v_{\rm full}(t)-v_M(t)\right]^2
},
\label{eq:phase_space_error}
\end{equation}
where $(x_{\rm full},v_{\rm full})$ denotes the trajectory obtained with the
full-history discrete GL reference and $(x_M,v_M)$ denotes the trajectory obtained by retaining only
the last $M+1$ GL coefficients. The corresponding memory horizon is $T_m=Mh$,
as defined in Eq.~\eqref{eq:memory_horizon}. To quantify the departure between
both trajectories, we define the coherence-loss time as
\begin{equation}
T_{\rm loss}
=
\inf\left\{t>0:\,E(t)>\varepsilon_{\rm loss}\right\},
\label{eq:Tloss}
\end{equation}
where $\varepsilon_{\rm loss}$ is a prescribed phase-space tolerance. If the
threshold is not reached during the simulated interval, we set
$T_{\rm loss}=t_{\rm final}$.

In this context, coherence means that the full-memory and finite-memory
trajectories remain indistinguishable within the prescribed tolerance
\(\varepsilon_{\rm loss}\). The value of \(\varepsilon_{\rm loss}\) therefore
fixes the resolution at which two trajectories are considered dynamically
distinguishable. The loss of coherence is an operational, tolerance-dependent
event: it occurs when the phase-space distance between the two trajectories
first exceeds the chosen threshold. It should not be interpreted as
synchronization loss between two coupled systems, nor as a universal stability
boundary. Rather, it measures the time interval over which the truncated-memory
realization provides a dynamically reliable surrogate of the full-memory GL
reference for the prescribed tolerance, discretization, and observation window.

We use the term coherence-loss time, rather than desynchronization time, because the comparison is not between two coupled oscillators but between two memory representations of the same system.

\begin{figure}[t]
\centering
\includegraphics[width=\textwidth]{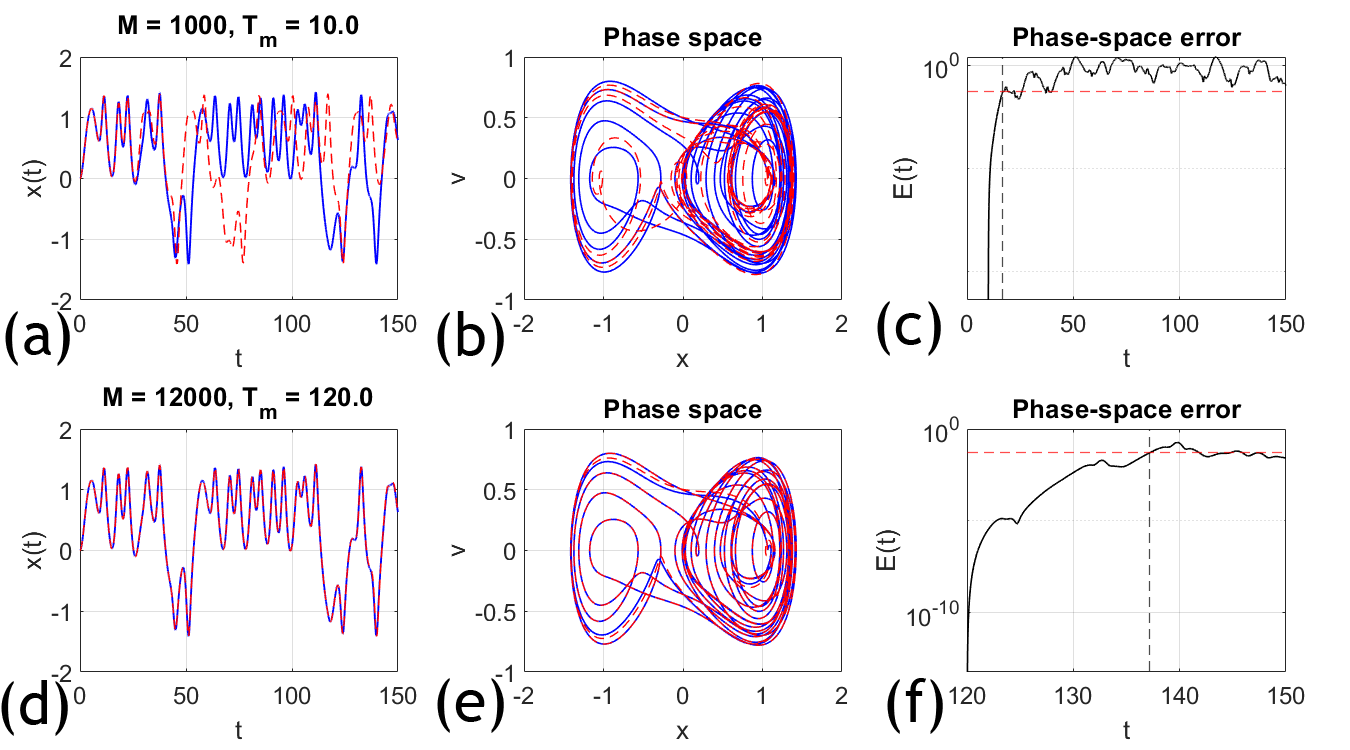}
\caption{
Direct comparison between the full-history discrete GL reference and its
truncated-memory counterpart for $q=0.70$, $f=0.30$, $\omega=1.20$, and
$h=0.01$. Panels (a)--(c) correspond to the shorter memory horizon
$M=1000$ ($T_m=10$), while panels (d)--(f) correspond to the longer memory
horizon $M=12000$ ($T_m=120$). Panels (a) and (d) show the time series,
panels (b) and (e) show the phase-space portraits, and panels (c) and (f)
show the phase-space error
$E(t)=\sqrt{[x_{\rm full}(t)-x_M(t)]^2+[v_{\rm full}(t)-v_M(t)]^2}$.
Blue solid curves denote the full-memory GL dynamics, while red dashed curves
denote the truncated-memory dynamics. For $M=1000$, the truncated trajectory
loses coherence with the full-memory reference at
$T_{\rm loss}=16.82$, using
$\varepsilon_{\rm loss}=5\times10^{-2}$. For $M=12000$, this loss is delayed
to $T_{\rm loss}=137.15$, showing that increasing the retained memory horizon
substantially extends the interval of dynamical agreement.
}
\label{fig:direct_comparison}
\end{figure}

Figure~\ref{fig:direct_comparison} shows a representative comparison for
$q=0.7$, $f=0.3$, $\omega=1.2$, and $h=0.01$. For the shorter memory horizon,
$M=1000$ ($T_m=10$), the truncated trajectory initially follows the full-memory
solution, but later departs from it. This loss of agreement is visible in the
time series, in the deformation of the phase portrait, and in the rapid growth
of the phase-space error. By contrast, for the longer memory horizon,
$M=12000$ ($T_m=120$), the truncated-memory dynamics remains close to the
full-memory reference over most of the observation interval.
Unless otherwise stated, the simulations in this section are performed with
time step \(h=0.01\) up to \(t_{\mathrm{final}}=150\), corresponding to
\(N=15000\) integration steps. The full-memory reference trajectory uses all
past states available at each time step, i.e. the GL sum extends from \(k=0\)
to \(k=n\) at time \(t_n\). By contrast, the truncated-memory trajectory retains
only the last \(M+1\) GL coefficients, from \(k=0\) to \(k=\min(n,M)\). Thus,
although \(M=12000\) corresponds to a long retained memory horizon,
\(T_m=Mh=120\), it is still a finite-memory approximation and does not use the
entire history over the full observation interval \(0\leq t\leq150\).
For the tolerance $\varepsilon_{\rm loss}=5\times10^{-2}$ used in this
comparison, the coherence-loss time increases from
$T_{\rm loss}=16.82$ for $M=1000$ to
$T_{\rm loss}=137.15$ for $M=12000$. Thus, increasing the memory horizon
from $T_m=10$ to $T_m=120$ extends the interval of dynamical agreement by
almost one order of magnitude.

This comparison highlights an important point: the truncated GL model should
not be regarded only as a cheaper numerical approximation of the full-memory
model. For insufficient memory horizons, the neglected tail of the fractional
kernel may be dynamically amplified, leading to a macroscopic departure of the
trajectory. In this sense, the parameter $T_m$ affects not only the computational cost of
the simulation, but also the temporal range over which the truncated system
remains dynamically coherent with the full-history discrete GL reference.

\subsection{Amplitude--phase diagnostic of coherence loss}

\begin{figure}[t]
\centering
\includegraphics[width=\textwidth]{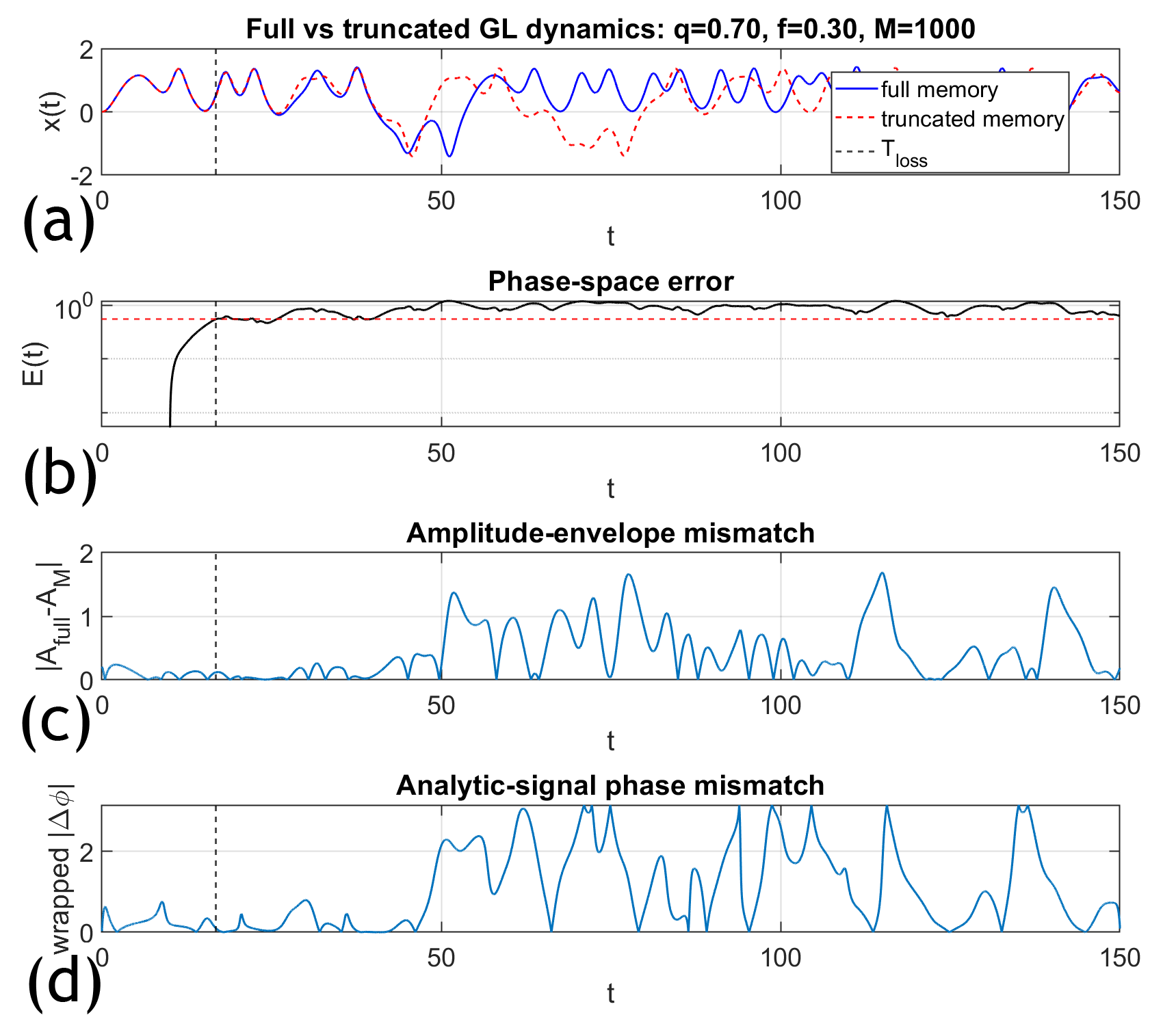}
\caption{
Amplitude--phase diagnostic of coherence loss for a representative sensitive
case, with $q=0.70$, $f=0.30$, and $M=1000$. Panel (a) shows the full-memory
and truncated-memory trajectories, panel (b) shows the phase-space error
$E(t)$ together with the loss threshold $\varepsilon_{\rm loss}$, panel (c)
shows the amplitude-envelope mismatch, and panel (d) shows the analytic-signal
phase mismatch. The vertical dashed line marks the coherence-loss time
$T_{\rm loss}$. After coherence is lost, both the amplitude and phase
diagnostics become appreciable, indicating that the departure between the full
and truncated dynamics is associated with a mixed amplitude--phase mismatch.
}
\label{fig:amp_phase_diagnostic}
\end{figure}

To further clarify the mechanism behind coherence loss, we separate the
discrepancy between the full-memory and truncated-memory trajectories into an
amplitude-envelope mismatch and an analytic-signal phase mismatch. The latter
should be understood as a phase-like diagnostic, since the response is not
necessarily a purely harmonic signal. Figure~\ref{fig:amp_phase_diagnostic}
shows that, once the phase-space error reaches the loss threshold at
$T_{\rm loss}$, both quantities become significant. This indicates that the
loss of coherence is not caused by a purely amplitude-related error nor by a
pure phase drift alone. Rather, in the dynamically sensitive regime shown here,
the separation between the full and truncated responses is associated with a
mixed amplitude--phase mismatch.

\section{Coherence loss and memory threshold in the \texorpdfstring{$(f,T_m)$}{(f,Tm)} plane}
\label{sec:coherence_loss}

\subsection{Phase-space error and coherence-loss time}

Using the phase-space error \(E(t)\) and the coherence-loss time
\(T_{\rm loss}\) defined in Eqs.~\eqref{eq:phase_space_error} and
\eqref{eq:Tloss}, we now quantify the effect of memory truncation in
parameter-space scans. For each parameter set, the full-history discrete GL reference and the
truncated-memory realization are integrated from the same initial data.
The retained memory horizon is \(T_m=Mh\), as defined in
Eq.~\eqref{eq:memory_horizon}. If the error threshold is not exceeded during
the simulation, we set \(T_{\rm loss}=t_{\rm final}\). In the computations
reported below we use \(\varepsilon_{\rm loss}=5\times10^{-2}\),
\(q=0.70\), \(\omega=1.20\), and \(t_{\rm final}=150\).

\subsection{Coherence-loss time as a predictability horizon}

The quantity $T_{\rm loss}$ provides an operational coherence horizon for the
truncated-memory approximation. It measures how long the trajectory generated
with a finite memory horizon remains dynamically coherent with the
full-history discrete GL reference trajectory. Thus, $T_{\rm loss}$ is not a synchronization
time in the usual sense, since the two trajectories do not correspond to two
coupled oscillators. Rather, it quantifies the time over which the truncated GL
operator provides a dynamically reliable surrogate of the full-memory GL
operator.

This distinction is important because the truncation error is not only a local
numerical defect in the evaluation of the fractional derivative. In nonlinear
regimes, the discarded memory tail acts as a deterministic perturbation that
can be amplified by the dynamics. Therefore, a small instantaneous difference
between the full and truncated memory terms may lead, after some time, to a
macroscopic separation between the two trajectories. In this sense,
$T_{\rm loss}$ provides an operational measure of the predictability of the
finite-memory approximation.

In the following, we use $T_{\rm loss}$ both at the level of individual
trajectories and in parameter-space scans. For a fixed tolerance
$\varepsilon_{\rm loss}$, larger values of $T_{\rm loss}$ indicate that the
truncated model remains coherent with the full-history discrete GL reference for a longer
time, while small values identify parameter regimes where the finite-memory
approximation loses predictive power rapidly.

\subsection{Coherence-loss map and critical memory threshold}

\begin{figure}[t]
\centering
\includegraphics[width=0.9\textwidth]{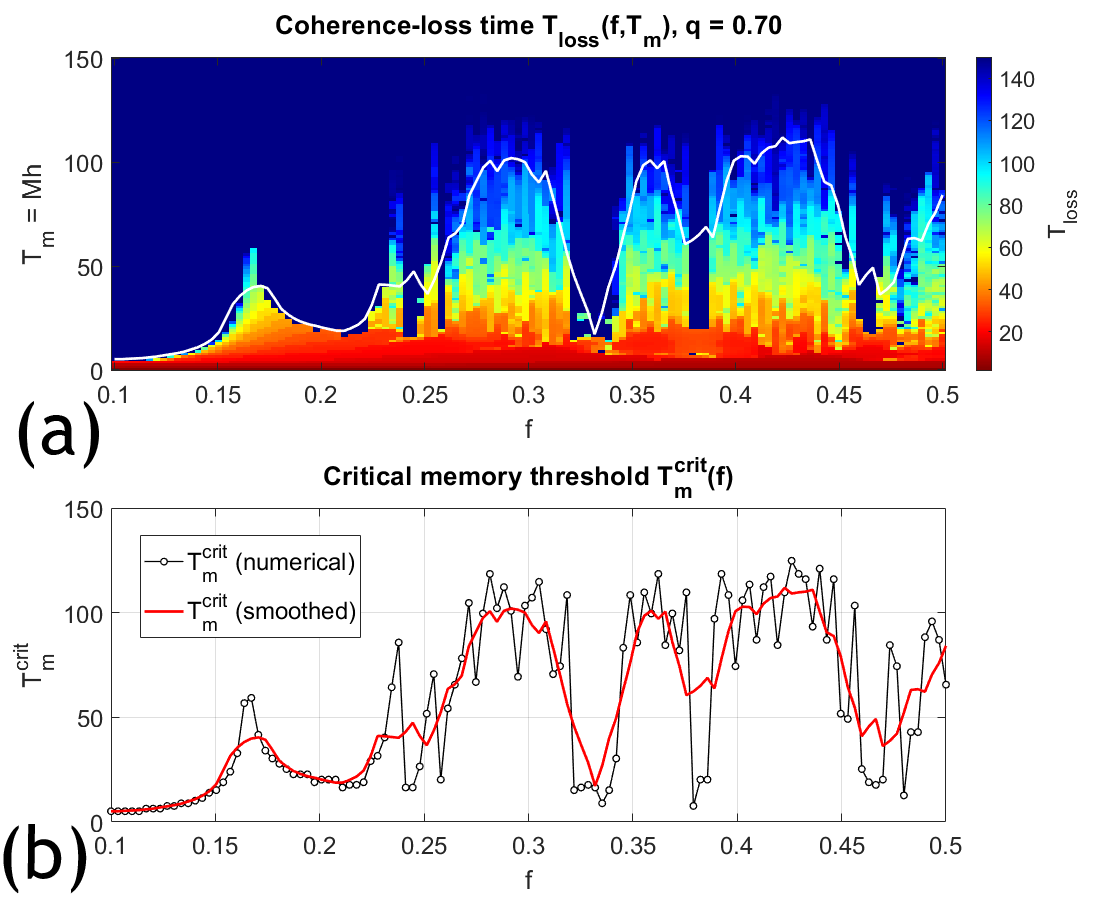}
\caption{
Coherence-loss map and critical memory threshold in the $(f,T_m)$ plane for
$q=0.70$, $\omega=1.20$, and $\varepsilon_{\rm loss}=5\times10^{-2}$.
Panel (a) shows the coherence-loss time $T_{\rm loss}(f,T_m)$, where
$T_m=Mh$ is the retained memory horizon. The white curve is a smoothed guide
to the eye obtained from the numerically extracted threshold values. Panel (b)
shows the extracted threshold
$T_m^{\rm crit}(f)=\min\{T_m:\,T_{\rm loss}(f,T_m)\geq T_{\rm target}\}$,
with $T_{\rm target}=0.95\,t_{\rm final}=142.5$. Black markers denote the
numerical threshold values, while the red curve is a smoothed guide to the eye
obtained from the same numerical points. The smoothed curves are included only
to highlight the trend and should not be interpreted as independent theoretical
predictions. The pronounced non-monotonic dependence on \(f\) indicates that the memory
horizon required to preserve coherence is associated with the underlying
dynamical regime, not only with the nominal truncation length. The scan uses 120 forcing values in $f\in[0.10,0.50]$ and $120$ memory depths
in $M\in[10,15000]$, with $T_m=Mh$ and $h=0.01$ and \(T_{\rm target}=0.95t_{\rm final}=142.5\). The superimposed curve
is a seven-point moving-average guide to the finite numerical threshold values;
it is included only to highlight the trend and does not represent an independent
theoretical fit.
}
\label{fig:Tmcrit}
\end{figure}

Figure~\ref{fig:Tmcrit} shows the resulting coherence-loss structure in the
$(f,T_m)$ plane. The upper panel displays the map of $T_{\rm loss}(f,T_m)$.
Dark blue regions correspond to parameter values for which the truncated
trajectory remains coherent with the full-history discrete GL reference over essentially
the whole observation interval. In contrast, red and yellow regions indicate
early loss of coherence. The map shows that the influence of memory truncation
is strongly dependent on the forcing amplitude. In particular, several bands
of enhanced sensitivity appear, where substantially larger memory horizons
are required to maintain agreement with the full-memory dynamics. The white
curve indicates a smoothed guide to the eye obtained from the numerically
extracted threshold values.

To summarize this information in a single quantity, we define a critical
memory threshold
\begin{equation}
T_m^{\rm crit}(f)
=
\min
\left\{
T_m:\,
T_{\rm loss}(f,T_m)\geq T_{\rm target}
\right\},
\label{eq:Tmcrit}
\end{equation}
where $T_{\rm target}$ is a prescribed target coherence time. In the present
case we take
\begin{equation}
T_{\rm target}=0.95\,t_{\rm final}=142.5,
\end{equation}
so that $T_m^{\rm crit}(f)$ represents the minimum retained memory horizon
needed for the truncated dynamics to remain coherent with the full-memory
reference for almost the entire simulated interval.

The lower panel of Fig.~\ref{fig:Tmcrit} shows the numerically extracted
threshold $T_m^{\rm crit}(f)$. The red curve is a smoothed interpolation of
these numerical threshold values and is included only as a guide to the eye;
it should not be interpreted as an independent theoretical prediction. The
resulting threshold is highly non-monotonic. For some forcing amplitudes, a
relatively short memory horizon is sufficient, whereas in other intervals the
required memory horizon increases by more than one order of magnitude. { This shows that the dynamical validity of the truncated GL realization is associated not only with the nominal size of \(M\), but also with the dynamical
regime selected by the forcing amplitude.

The operator-level estimate derived in Sec.~II can be combined with a
finite-time amplification assumption in order to provide a more explicit
rationale for the observed memory-cliff structure. Let
\(\mathbf{z}(t)=(x(t),v(t))\) denote the full-history discrete GL trajectory
and let \(\mathbf{z}*M(t)\) denote the corresponding truncated-memory
trajectory. Suppose that, over the observation interval and in the region
explored by both trajectories, perturbations satisfy a finite-time propagation
bound of the form
\begin{equation}
|G(t,s)|
\leq
C_G e^{\Lambda(t-s)},
\qquad
0\leq s\leq t,
\label{eq:finite_time_propagation_bound}
\end{equation}
where (G(t,s)) is the associated perturbation-propagation operator and
\(\Lambda\) is an effective finite-time amplification rate. Combining this
assumption with the tail estimate in Eq.~\eqref{eq:tail_bound_Tm} gives the
conditional trajectory estimate
\begin{equation}
E(t)
\lesssim
C*{\rm dyn}(q,X),T_m^{-q}
\frac{e^{\Lambda t}-1}{\Lambda},
\qquad
\Lambda>0,
\label{eq:conditional_error_bound}
\end{equation}
with the limiting form
\begin{equation}
E(t)
\lesssim
C_{\rm dyn}(q,X),T_m^{-q}t,
\qquad
\Lambda=0.
\label{eq:conditional_error_bound_zero}
\end{equation}

Equations~\eqref{eq:conditional_error_bound} and
\eqref{eq:conditional_error_bound_zero} are conditional finite-time estimates,
not a global stability theorem for the nonlinear fractional Duffing system.
They make explicit the two ingredients underlying the observed memory cliffs:
the discarded GL tail decreases algebraically with the retained memory horizon,
whereas its effect may be amplified over finite times in dynamically sensitive
parameter regions.

The abrupt variations of $T_m^{\rm crit}$ observed in the numerical maps can
therefore be interpreted phenomenologically as the result of these two
competing effects. Memory truncation introduces a deterministic perturbation
associated with the discarded GL tail,

\begin{equation}
\delta_M(t_n)
=
D_h^{q,\mathrm{GL}}x_n
-
D_h^{q,\mathrm{GL},M}x_n.
\end{equation}
If this perturbation is amplified over a finite time interval with an effective
rate $\lambda_{\rm eff}(q,f)$, one may write, only as a local heuristic
description,
\begin{equation}
E(t)\sim E_0(M,q)\,e^{\lambda_{\rm eff}(q,f)t},
\end{equation}
where $E_0(M,q)$ represents an effective size of the truncation-induced
perturbation. Since the GL coefficients decay algebraically,
$|w_k(q)|\sim k^{-q-1}$, the discarded tail suggests the approximate scaling
\begin{equation}
E_0(M,q)\sim C(q)M^{-q}.
\end{equation}
Combining these two heuristic ingredients gives
\begin{equation}
T_{\rm loss}
\sim
\frac{1}{\lambda_{\rm eff}(q,f)}
\left[
\log\left(\frac{\varepsilon_{\rm loss}}{C(q)}\right)
+
q\log M
\right],
\label{eq:Tloss_scaling_heuristic}
\end{equation}
or, equivalently, for a prescribed target coherence time,
\begin{equation}
M_{\rm crit}
\sim
\left[
\frac{C(q)}{\varepsilon_{\rm loss}}
e^{\lambda_{\rm eff}(q,f)T_{\rm target}}
\right]^{1/q}.
\label{eq:Mcrit_scaling_heuristic}
\end{equation}
The corresponding critical memory horizon is
$T_m^{\rm crit}=hM_{\rm crit}$.

These expressions are not fitted laws and are not used to predict the
numerically extracted values of $T_{\rm loss}$ or $T_m^{\rm crit}$. They are
introduced only as a phenomenological rationale for the observed memory cliffs:
an algebraically small discarded GL tail can become dynamically relevant when
its effect is amplified by a sensitive nonlinear response. In particular,
$C(q)$ and $\lambda_{\rm eff}(q,f)$ should be regarded as effective descriptive
quantities rather than independently calibrated model parameters.

Thus, the abrupt memory-threshold transition observed in the numerical maps is
consistent with a dynamical amplification mechanism: the discarded GL
tail acts as a deterministic perturbation whose effect is magnified by the
nonlinear sensitivity of the underlying Duffing response. We refer to this type
of sharp transition as a ``memory cliff'', namely a narrow region in parameter
space where a relatively small change in the retained memory horizon produces a
large change in the coherence-loss time $T_{\rm loss}$ or in the extracted
critical threshold $T_m^{\rm crit}$.}

As a complementary diagnostic, we computed a finite-time sensitivity proxy for
the full-history discrete GL reference dynamics as a function of the forcing amplitude. The
regions where this proxy becomes large coincide qualitatively with intervals
where the coherence-loss time decreases and the critical memory horizon
increases. This supports the interpretation that the discarded GL tail acts as
a small deterministic perturbation whose effect is amplified in dynamically
sensitive regimes. We emphasize that this quantity is used only as a finite-time
sensitivity proxy, not as a rigorous Lyapunov exponent for the fractional
memory dynamics.
{
A direct comparison between this finite-time sensitivity indicator and rigorous
Lyapunov exponents is beyond the scope of the present work and is left for
future study. Likewise, Eqs.~\eqref{eq:Tloss_scaling_heuristic}--
\eqref{eq:Mcrit_scaling_heuristic} should be interpreted exclusively as a
phenomenological scaling argument, not as a quantitative predictive theory.
A systematic calibration of $C(q)$ and $\lambda_{\rm eff}(q,f)$, together with
a direct comparison between the resulting estimates and the numerically
extracted values of $T_{\rm loss}$ and $T_m^{\rm crit}$, is left for future
work.
}

\section{Complementary analysis in the \texorpdfstring{$(q,T_m)$}{(q,Tm)} plane}
\label{sec:q_Tm_analysis}

The previous section showed that the memory horizon required to preserve
coherence depends strongly on the forcing amplitude. We now perform a
complementary analysis by fixing the forcing amplitude and varying the
fractional order. This allows us to assess how the fractional order of the GL operator affects the sensitivity of the dynamics to memory truncation.

We fix $f=0.30$ and $\omega=1.20$, and compute the coherence-loss time
$T_{\rm loss}$ in the $(q,T_m)$ plane using the same phase-space error
defined in Eq.~\eqref{eq:phase_space_error} and the same tolerance
$\varepsilon_{\rm loss}=5\times10^{-2}$. As before, when the error does not
exceed the tolerance during the simulated interval, we set
$T_{\rm loss}=t_{\rm final}$.

\begin{figure}[t]
\centering
\includegraphics[width=0.9\textwidth]{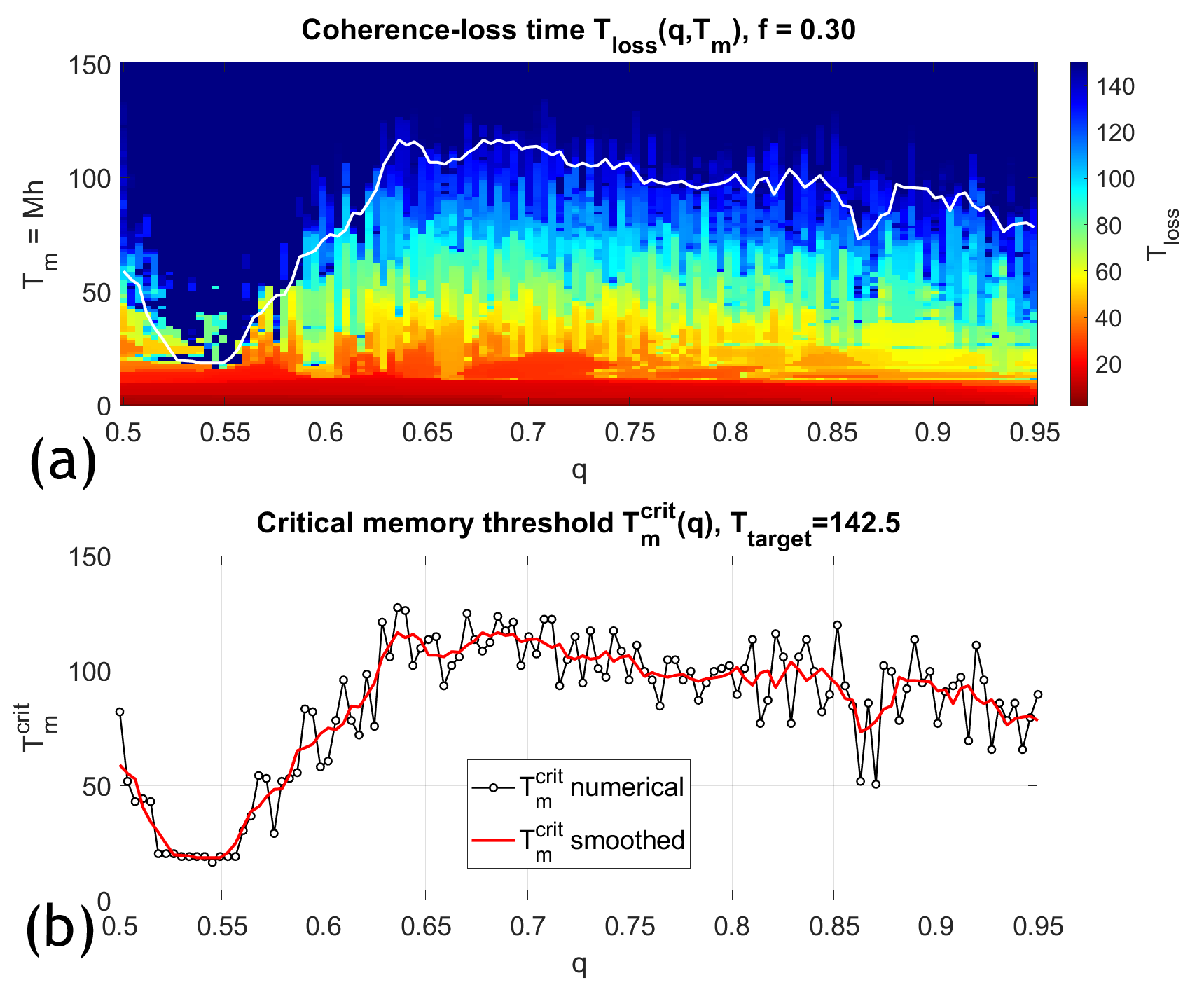}
\caption{
Complementary coherence-loss analysis in the $(q,T_m)$ plane for fixed
forcing amplitude $f=0.30$ and $\omega=1.20$.
Panel (a) shows the coherence-loss time $T_{\rm loss}(q,T_m)$, computed from
the phase-space error between the full-history discrete GL reference and its
truncated-memory counterpart. The white curve is a smoothed guide to the eye
obtained from the numerically extracted threshold values.
Panel (b) shows the extracted threshold
$T_m^{\rm crit}(q)=\min\{T_m:\,T_{\rm loss}(q,T_m)\geq T_{\rm target}\}$,
with $T_{\rm target}=142.5$. Black markers denote the numerical threshold
values, while the red curve is a smoothed guide to the eye obtained from the
same numerical points. These smoothed curves are included only to highlight
the trend and should not be interpreted as independent theoretical
predictions. The dependence on $q$ is strongly non-monotonic: after a
low-threshold region near $q\simeq0.53$--$0.56$, the required memory horizon
increases sharply and remains large over most of the intermediate- and
large-$q$ range. The scan uses $120$ fractional orders in $q\in[0.50,0.95]$ and $120$ memory
depths in $M\in[10,15000]$, with $T_m=Mh$ and $h=0.01$ and \(T_{\rm target}=0.95t_{\rm final}=142.5\). The
superimposed curve is a five-point moving-average guide to the finite numerical
threshold values; it is included only to highlight the trend and does not
represent an independent theoretical fit.
}
\label{fig:qTm_loss}
\end{figure}

Figure~\ref{fig:qTm_loss} shows the resulting map of
$T_{\rm loss}(q,T_m)$ together with the critical memory threshold
$T_m^{\rm crit}(q)$, computed using the same target coherence time
$T_{\rm target}=0.95\,t_{\rm final}=142.5$. The results reveal a strongly
non-monotonic dependence on the fractional order. In particular, relatively
short memory horizons are sufficient in a narrow region around
$q\simeq 0.53$--$0.56$, whereas the required memory horizon increases sharply
as $q$ approaches the intermediate range $q\simeq 0.6$--$0.7$.

For larger values of $q$, the threshold remains high over a broad interval,
although with significant fluctuations. This behavior indicates that the effect of memory truncation is not determined solely by the algebraic decay rate of the GL coefficients. Instead, the dynamical regime selected by the
fractional order also plays an important role in determining how strongly the
neglected memory tail is amplified by the nonlinear forced dynamics.

\begin{figure*}[t]
\centering
\includegraphics[width=\textwidth]{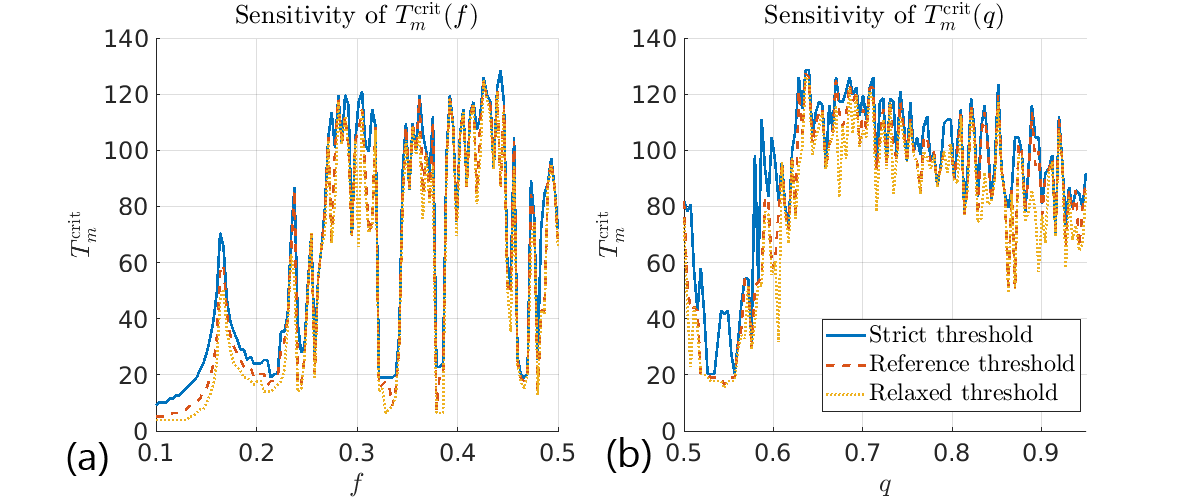}
\caption{{Threshold-sensitivity check for the operational critical memory horizon.
Panel (a) shows \(T_m^{\rm crit}(f)\) for \(q=0.70\) and \(\omega=1.20\),
whereas panel (b) shows \(T_m^{\rm crit}(q)\) for \(f=0.30\) and
\(\omega=1.20\). In both panels, the strict, reference, and relaxed curves
correspond, respectively, to \(\varepsilon_{\rm loss}=0.025\),
\(\varepsilon_{\rm loss}=0.05\), and \(\varepsilon_{\rm loss}=0.10\). The
remaining numerical settings are those of the corresponding main scans,
including \(h=0.01\), \(t_{\rm final}=150\), and
\(T_{\rm target}=0.95t_{\rm final}=142.5\). Although the absolute values of
\(T_m^{\rm crit}\) vary with the prescribed tolerance, the main non-monotonic
structures and high-sensitivity regions persist across the three thresholds.}}
\label{fig:threshold_sensitivity}
\end{figure*}

{
To assess the dependence of the extracted threshold on the operational
coherence criterion, we repeated representative one-dimensional scans using
three values of $\varepsilon_{\rm loss}$. Figure~\ref{fig:threshold_sensitivity}
shows that tightening or relaxing the threshold shifts the absolute value of
$T_m^{\rm crit}$, as expected. However, the principal non-monotonic features
of both $T_m^{\rm crit}(f)$ and $T_m^{\rm crit}(q)$ remain visible across the
three tolerance levels. These results support the interpretation of the
memory-cliff curves as tolerance-dependent diagnostic boundaries rather than
universal dynamical thresholds: their precise location is operational, whereas
the associated parameter regions of enhanced memory sensitivity persist under
moderate changes of the coherence tolerance.
}

\section{Local spectral interpretation of the truncated dynamics}
\label{sec:local_spectral}

We now complement the coherence-loss analysis with a local spectral
interpretation of the truncated GL dynamics. The purpose of this section is
not to derive an exact global bifurcation condition for the forced nonlinear
system, but to test whether the characteristic equation associated with the
truncated memory kernel provides a useful indicator of the transition region
observed in the late-time dynamics. For each value of the forcing amplitude, we integrate the truncated GL system
with fixed memory horizon $T_m=Mh$ and extract the local maxima of $x(t)$ after
discarding the transient. In this section we use a longer integration time,
$t_{\rm final}=300$, in order to improve the sampling of the asymptotic
response. The late-time mean
\begin{equation}
x^*(f)=\langle x(t)\rangle_{\rm tail}
\end{equation}
is used as an effective reference state for a local linearization. The
corresponding local stiffness is
\begin{equation}
b_0(f)=\alpha+3\beta [x^*(f)]^2 .
\end{equation}
The characteristic equation associated with the truncated GL kernel is then
written as
\begin{equation}
\Delta_{\rm GL}(\lambda;f)
=
\lambda^2+b_0(f)+c h^{-q}K_{\rm GL}(\lambda),
\end{equation}
where
\begin{equation}
K_{\rm GL}(\lambda)
=
\sum_{k=0}^{M}w_k e^{-\lambda kh}.
\end{equation}
We also compare this expression with the characteristic equation obtained by
replacing the truncated GL kernel with its positive-delay exponential
representation,
\begin{equation}
K_{\rm pos}(\lambda)
=
A_0+\sum_{j=1}^{r}A_j e^{-\lambda \tau_j},
\qquad \tau_j>0,
\end{equation}
leading to
\begin{equation}
\Delta_{\rm pos}(\lambda;f)
=
\lambda^2+b_0(f)+c h^{-q}K_{\rm pos}(\lambda).
\end{equation}

As a local Hopf-type spectral indicator, we compute
\begin{equation}
R(f)=\min_{\omega\geq0}|\Delta(i\omega;f)|.
\end{equation}
Small values of $R(f)$ indicate that the local characteristic equation is close
to satisfying a purely imaginary-root condition. This criterion should be
understood as a local spectral indicator, rather than as an exact bifurcation
condition for the full forced nonlinear dynamics.

\begin{figure}[t]
\centering
\includegraphics[width=\textwidth]{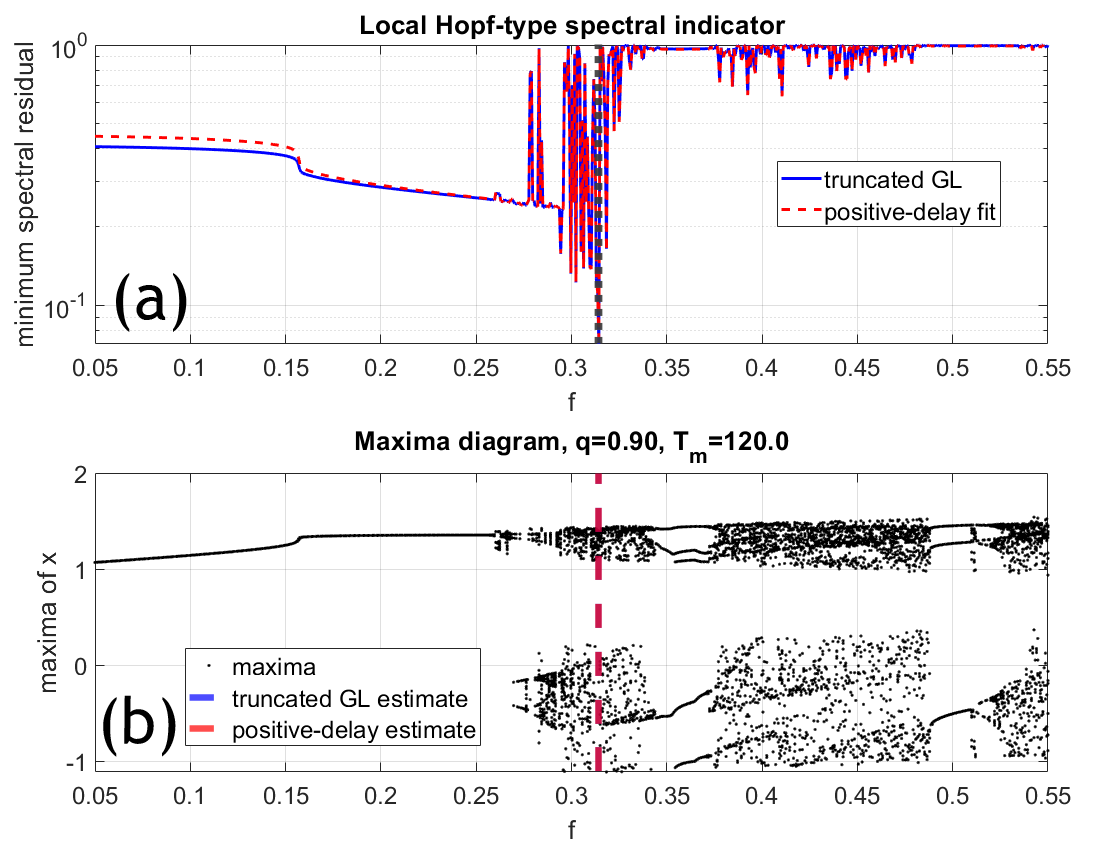}
\caption{{
Local spectral interpretation of the truncated GL dynamics for $q=0.90$ and
$T_m=120$. Panel (a) shows the minimum spectral residual
$R(f)=\min_{\omega\geq0}|\Delta(i\omega;f)|$ computed from the characteristic
equation of the truncated GL kernel and from its positive-delay exponential
approximation. Panel (b) shows the corresponding diagram of late-time maxima
of $x(t)$ as the forcing amplitude is varied. The vertical lines mark the
values of $f$ at which the spectral residuals attain their minima. These lines
should be interpreted as local spectral estimates of the transition region,
not as exact bifurcation points of the forced nonlinear system. For this representative case, the truncated GL characteristic equation and the
positive-delay approximation yield essentially the same local spectral estimate,
located near $f\simeq0.298$. This agreement should be understood as evidence
that the positive-delay representation reproduces the local spectral structure
of the truncated kernel over the selected fitting window, rather than as a
general bifurcation criterion for the full nonlinear dynamics.}}
\label{fig:local_spectral_maxima}
\end{figure}

{
Figure~\ref{fig:local_spectral_maxima} provides representative evidence that
the minimum of the local spectral residual lies close to the transition region
observed in the maxima diagram. For the case shown, the estimates obtained from
the truncated GL characteristic equation and from the positive-delay
representation are almost indistinguishable at the scale of the figure. Both
approaches yield $f_H\simeq0.298$, with only a small difference in the
associated spectral frequency.

The vertical lines should not be interpreted as exact bifurcation points of the
forced nonlinear system. They mark the minima of a local spectral residual
computed from an effective linearization around the late-time mean state, and
therefore provide a diagnostic associated with the observed transition region
rather than a precise criterion for the onset of the changes seen in the maxima
diagram. For the representative case analyzed here, the truncated GL kernel and
its positive-delay representation lead to almost identical spectral estimates.
This agreement indicates that the delay representation reproduces the local
spectral structure of the truncated memory operator over the selected spectral
window.

It is important to stress that this analysis does not imply that the truncated
GL system is globally equivalent to a finite-delay equation, nor that the local
spectral residual constitutes a general bifurcation criterion for the forced
nonlinear dynamics. Rather, the positive-delay representation provides a local
spectral surrogate for the truncated kernel. Its usefulness is illustrated here
through one representative comparison with the transition structure observed in
the nonlinear maxima diagram.
}

\section{Positive-delay representation and memory-kernel complexity}
\label{sec:delay_complexity}

The local approximation developed in Sec.~\ref{sec:local_spectral} shows that
the truncated GL kernel admits a delay-type spectral interpretation. In its
simplest form, this interpretation can be written as the sum of an
instantaneous contribution and a single delayed exponential term. However,
this minimal representation is only a low-order local surrogate and does not,
in general, capture the full spectral structure of the truncated finite-memory
kernel over the selected spectral window. This naturally raises a more general
question: can the same truncated kernel be represented accurately by a finite
number of positive-delay exponential modes, and how many such modes are
required?

We address this question by introducing a positive-delay representation of the
truncated GL kernel and by defining the minimum delay complexity \(r_{\min}\).
We now analyze the structure of this representation more systematically in the
parameter plane spanned by the fractional order \(q\) and the memory horizon
\(T_m\).

For fixed $q$ and truncation length $M$, the truncated GL memory term defines
the kernel
\begin{equation}
K_{\rm GL}(\lambda;q,M)
=
\sum_{k=0}^{M} w_k(q)e^{-\lambda kh}.
\end{equation}
We approximate this kernel by a finite positive-delay exponential
representation of the form
\begin{equation}
K_r(\lambda)
=
A_0+\sum_{j=1}^{r}A_j e^{-\lambda \tau_j},
\qquad \tau_j>0 .
\end{equation}

The coefficients \(A_0\), \(A_j\), and the delays \(\tau_j\) are obtained by
minimizing a relative error over a local spectral domain
\(\lambda=\sigma+i\omega\). To avoid confusion with the trajectory error
\(E(t)\), we denote this spectral kernel-fitting error by
\(\mathcal{E}_r(q,M)\), 
\begin{equation}
\mathcal{E}_r(q,M)
=
\frac{
\left\|K_r(\lambda)-K_{\rm GL}(\lambda;q,M)\right\|
}{
\left\|K_{\rm GL}(\lambda;q,M)\right\|
}.
\label{eq:kernel_fit_error}
\end{equation}
{
In the numerical scan, the error is evaluated over the rectangular domain
\(0\leq \sigma\leq0.08\) and \(0\leq\omega\leq0.80\), sampled on a uniform
\(24\times24\) grid. The minimization is performed using MATLAB's
\texttt{lsqnonlin}, applied jointly to the real and imaginary parts of the
kernel over the selected \(\lambda\)-grid. The amplitudes \(A_0\) and \(A_j\)
are left unconstrained, whereas positivity of the delays is enforced through
the parametrization \(\tau_j=\exp(\theta_j)\).

Since the optimization problem is non-convex, each fit is initialized from
several random starting points. For each value of \(r\), the initial delays are
sampled logarithmically in the interval \([h,T_m]\), the offset \(A_0\) is
initialized near the mean value of the exact kernel with a small random
perturbation, and the amplitudes \(A_j\) are initialized from random values.
In the scans reported below, we use 12 random starts for each value of \(r\)
and retain the solution with the smallest relative spectral error.

A limited robustness check with 24 and 48 starts at representative parameter
points showed that the broad complexity classification is preserved, whereas
the extracted value of \(r_{\min}\) may decrease by one or more modes near
difficult fitting regions. Thus, the map obtained with 12 starts should be
interpreted as an operational, and potentially conservative, estimate of the
minimum positive-delay complexity. If no fit with
\(r\leq r_{\max}=9\) reaches the prescribed tolerance
\(\varepsilon_{\rm fit}=10^{-2}\), the corresponding parameter point is
recorded as unresolved.

}

We define the minimum acceptable number of positive delays as
\begin{equation}
r_{\min}(q,M)
=
\min \left\{ r:\, \mathcal{E}_r(q,M)<\varepsilon_{\rm fit}\right\},
\end{equation}
with $\varepsilon_{\rm fit}=10^{-2}$.

The mathematical meaning of \(r_{\min}\) is spectral and operational. The
truncated GL operator already contains a finite set of delayed samples
\(x(t-kh)\), \(k=0,\ldots,M\), with algebraically weighted coefficients inherited
from the fractional derivative. In this sense, memory truncation converts the
infinite-history fractional operator into a finite distributed-memory kernel
with delay-type structure. However, the individual lags \(kh\) should not be
interpreted as physical delays imposed on the oscillator. They are numerical
memory locations generated by the GL discretization.

The positive-delay representation provides a compressed local spectral
surrogate of this finite-memory kernel. A small value of \(r_{\min}\) means that
the action of the truncated GL kernel over the selected spectral window can be
reproduced by only a few causal exponential delay modes. A large value of
\(r_{\min}\) indicates that the finite-memory kernel has a richer distributed
structure and cannot be reduced, at the prescribed tolerance, to a small number
of effective delayed contributions. Thus, \(r_{\min}\) should be interpreted as
an operational measure of the causal delay-like complexity, or compressibility,
of the truncated memory kernel, not as the number of physical delays present in
the original Duffing oscillator.

This connection with delay-induced dynamics is therefore not based on the
presence of an externally imposed feedback delay, but on the exponential
structure of the characteristic equation generated by the finite GL memory
kernel. The truncated memory acts as a distributed collection of past
contributions, and the positive-delay approximation quantifies how many causal
delay-like modes are needed to reproduce its local spectral effect. Below the
threshold \(r_{\min}\), the positive-delay representation is not sufficiently
accurate in the selected spectral window; at or above it, the finite-memory GL
kernel admits a causal delay-type surrogate at the chosen accuracy level.

{
The value of \(r_{\min}\) is necessarily tolerance-dependent and depends on
the selected spectral window, fitting protocol, and maximum number of tested
exponential modes. To assess the dependence on the tolerance, we recalculated
\(r_{\min}\) from the same optimized error curves using
\(\varepsilon_{\rm fit}=5\times10^{-3}\), \(10^{-2}\), and
\(5\times10^{-2}\), without performing any additional fitting. As expected,
stricter tolerances increase the required number of positive-delay modes and
may leave some points unresolved within the tested range \(r\leq9\), whereas
looser tolerances reduce the apparent delay complexity.

The value \(r_{\max}=9\) is therefore not proposed as a theoretical upper
bound on the delay complexity of the truncated kernel. It is an operational
maximum selected for the present parameter scan, sufficient to resolve the
majority of the sampled points at the reference tolerance while keeping the
nonlinear fitting procedure computationally tractable. Points that remain
unresolved at \(r=9\) are explicitly retained as unresolved rather than being
assigned an artificial finite value. Thus, \(r_{\min}\) should not be
interpreted as an intrinsic invariant of the kernel, but as an operational
complexity measure at a prescribed spectral accuracy and within the tested
range of positive-delay modes.
}

\begin{figure}[t]
\centering
\includegraphics[width=\textwidth]{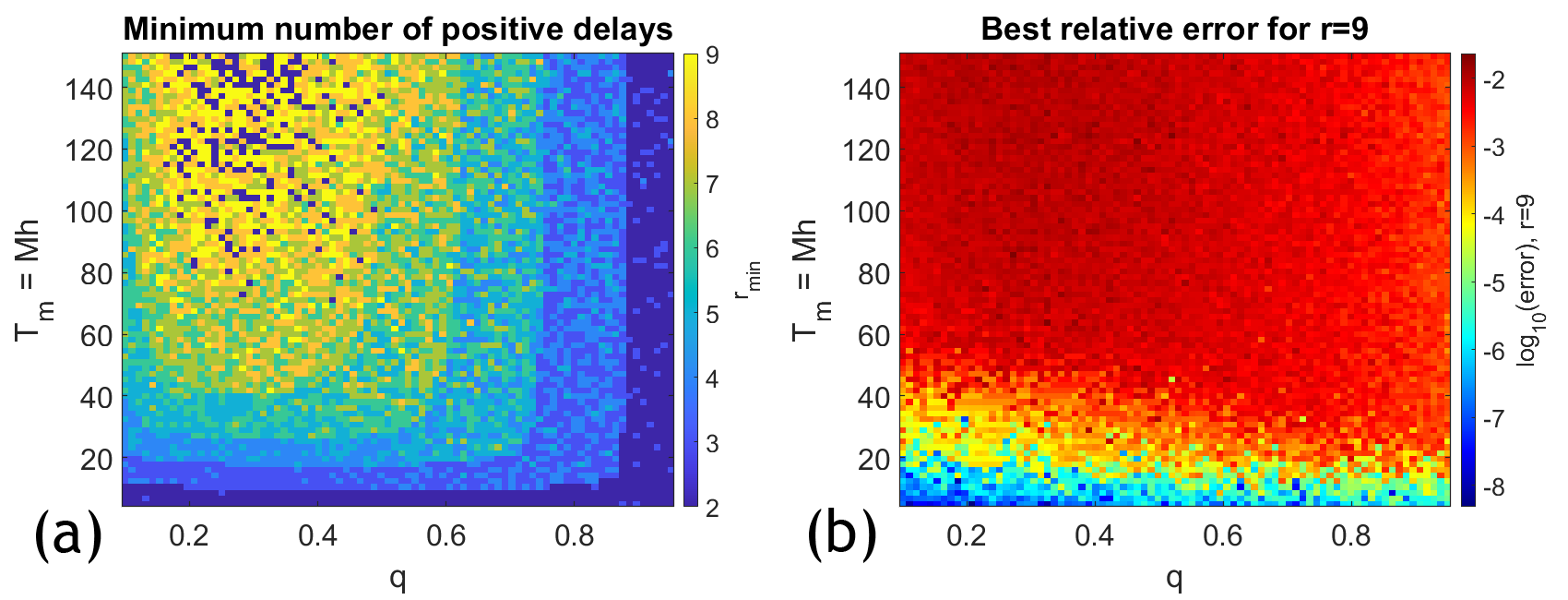}
\caption{
Positive-delay complexity of the truncated GL kernel in the $(q,T_m)$ plane.
Panel (a) shows the minimum number of positive-delay exponentials $r_{\min}$
required to approximate the truncated GL kernel with relative error below
$\varepsilon_{\rm fit}=10^{-2}$. Panel (b) shows the best relative error
obtained for the largest tested number of exponentials, $r=9$. The scan shows
that the effective delay complexity depends jointly on the fractional order
$q$ and on the memory horizon $T_m=Mh$. Short memory horizons are generally
well represented with few modes, while long memory horizons, especially for
low-to-intermediate fractional orders, may require several positive-delay
exponentials to reproduce the local spectral structure of the truncated GL
kernel.
}
\label{fig:delay_complexity_qTm}
\end{figure}

Figure~\ref{fig:delay_complexity_qTm} shows the resulting delay-complexity map
in the $(q,T_m)$ plane. For short memory horizons, only a small number of
positive delays is required. As the retained memory horizon increases, the
effective complexity generally grows and the kernel requires more exponential
modes to reach the same tolerance. The dependence on $q$ is strongly
nonuniform. For long memory horizons, low-to-intermediate fractional orders
typically require several positive-delay modes, whereas the large-$q$ region
appears more compressible over a broad portion of the parameter plane.

The right panel of Fig.~\ref{fig:delay_complexity_qTm} shows the best relative
error obtained for the largest tested number of exponentials, $r=9$. This map
provides a complementary diagnostic of the quality of the representation at
the maximum tested complexity. The smallest errors are obtained for short
memory horizons, whereas larger horizons generally lead to errors closer to
the prescribed tolerance. Together, the two panels show that the truncated GL
kernel is not generally equivalent to a single effective delay. Instead, its
local spectral action may require a small but nontrivial set of positive-delay
modes, whose number depends jointly on $q$ and $T_m$.

\section{Scope, limitations, and relation to existing approaches}

The results presented above should be understood within a precise scope. The main
objective of this work is not to propose a new general-purpose approximation of fractional
derivatives, but to analyze the dynamical consequences of using a finite-memory
Grünwald--Letnikov operator in a nonlinear forced oscillator. In this sense, the truncated
GL model is treated as a modified dynamical system in its own right, rather than only as a
computationally cheaper version of the full-memory fractional model.

This viewpoint differs from standard frequency-domain approximations of
fractional operators, such as rational, Padé-type, or Oustaloup-like
representations, whose purpose is usually to approximate the continuous
operator over a prescribed frequency band. Here, by contrast, the object being
approximated is the discrete truncated GL kernel actually used in the time
integration. Consequently, the positive-delay representation introduced in
this work should be interpreted as a local spectral surrogate of the
finite-memory kernel, not as a universal replacement for the fractional
derivative.

Recent exponential-sum, adaptive-memory, and kernel-compression approaches
reduce the cost of fractional-memory calculations by replacing a long history
convolution with a smaller set of recursively updated auxiliary contributions
\cite{Schadle2006,Jiang2017,Baffet2017,MacDonald2015,Rekanos2014}. In
particular, some approaches approximate the coefficients entering a
Gr\"unwald--Letnikov realization by sums of decaying exponentials, thereby
reducing storage requirements and the cost of evaluating the discrete memory
convolution.

The present construction is related to these approaches in its use of
exponential representations, but differs in both object and purpose. Here the
fitted object is the already truncated finite GL kernel, rather than the
continuous power-law kernel or the full discrete convolution considered for
computational acceleration. Moreover, the positive-delay representation is not
introduced as a fast solver. It is used as a local spectral surrogate in order
to quantify the effective positive-delay complexity of the finite-memory
operator and to relate that complexity to the dynamical consequences of memory
truncation.

It is useful to distinguish three levels of interpretation in the present
work. First, the truncated GL operator and its finite kernel are exact objects
at the discrete numerical level. Second, the characteristic equation, the
single-delay expansion, and the positive-delay representation provide local
spectral approximations, valid only in the selected spectral domain. Third,
the coherence-loss time, critical memory horizon, memory cliffs, and
delay-complexity maps are operational diagnostics: they quantify the dynamical
and spectral consequences of truncation for the chosen tolerance, observation
time, discretization, spectral domain, and parameter ranges. Thus, the results
should not be read as universal invariants of fractional dynamics, but as
reproducible measures of how a finite-memory GL implementation modifies the
effective dynamics.

Several limitations follow from this interpretation. First, the delay-type representation is
local in the spectral variable and is constructed over a finite window of $\lambda=\sigma+i\omega$.
Therefore, it does not imply a global equivalence between the truncated GL dynamics and a
finite-delay differential equation. Second, the Hopf-type criterion based on the minimum
spectral residual is only a local indicator of transition regions. It is not an exact bifurcation
condition for the full nonlinear periodically forced system. Third, the extracted delay
complexity $r_{\min}$ depends on the prescribed tolerance, on the selected spectral window,
and on the maximum number of exponentials allowed in the fit. It should therefore be
regarded as an operational measure of kernel complexity, not as an intrinsic invariant of
the fractional operator.

The numerical thresholds reported in the coherence-loss maps have a similar status. The
critical memory horizon $T_m^{\rm crit}$ depends on the error tolerance, on the observation
time, and on the dynamical regime selected by the forcing amplitude and the fractional
order. Thus, the memory cliffs identified in the parameter scans should not be interpreted
as universal boundaries. Rather, they quantify where, for the chosen diagnostics, the
discarded GL tail is dynamically amplified enough to destroy coherence with the
full-history discrete GL reference trajectory.

These limitations are also what define the contribution of the present work. 

 The framework developed here is not restricted to the symmetric forced Duffing
oscillator considered as a representative case. Its application to another
nonlinear oscillator requires four basic ingredients: a full-history reference
memory realization, a finite-memory counterpart obtained by truncating the same
kernel, a trajectory-level metric for quantifying their separation, and a local
spectral or linearized representation of the truncated memory operator over a
selected region of interest.

Under these conditions, the coherence-loss time and critical memory horizon can
be defined for a broad class of systems with fractional memory, including
asymmetric Duffing oscillators, multi-well oscillators, Helmholtz-type systems,
van der Pol-type oscillators, excitable systems such as FitzHugh--Nagumo
models, and coupled nonlinear oscillators. The corresponding delay-complexity
analysis can also be transferred whenever the truncated memory kernel admits a
finite exponential representation in the local spectral domain considered.

Preliminary exploratory calculations on a memory-focused fractional logistic
map provide supporting evidence beyond continuous-time oscillators. In that
discrete setting, finite-memory truncation also produces strongly
parameter-dependent retained-memory requirements, localized memory-cliff
regions, and a nontrivial relation between trajectory-level memory demand and
positive-delay kernel complexity. Because the diagnostics and numerical setting
are not identical to those used here, this comparison is presented only as
preliminary supporting evidence; a systematic cross-model analysis is left for
future work.

The quantitative maps of \(T_{\rm loss}\), \(T_m^{\rm crit}\), and
\(r_{\min}\) will generally depend on the specific nonlinear vector field,
forcing protocol, fractional order, and dynamical regime. Therefore, the
present results should not be interpreted as universal numerical thresholds
for all fractional oscillators. Rather, the generalizable contribution is the
diagnostic framework itself: finite-memory truncation can be assessed through
the combination of operator-tail control, trajectory-level coherence, and
local positive-delay compressibility.

The analysis
shows that memory truncation is not a neutral numerical operation in nonlinear fractional
systems. It modifies the effective memory kernel and may produce regime-dependent loss
of coherence. At the same time, the truncated GL kernel possesses a delay-like local
spectral structure that can be quantified through positive-delay exponential
representations. The present framework therefore provides a bridge between fractional
memory, finite-memory numerical implementations, and delay-type dynamical systems,
while making explicit that this bridge is local, diagnostic, and regime-dependent.

\section{Conclusions}
\label{sec:conclusions}

In this work we have investigated how truncating the Gr\"unwald--Letnikov memory
affects the dynamics of a forced Duffing oscillator with fractional damping.
The analysis was organized around two complementary questions: how long a
truncated-memory trajectory remains dynamically coherent with the full-history discrete GL reference, and how complex the truncated memory kernel is when represented in
terms of effective positive delays.

The first main finding is that memory truncation is not a neutral numerical
operation in nonlinear fractional systems. The direct comparison between the full-history discrete GL reference and the truncated-memory dynamics showed that the neglected tail of
the GL kernel can be dynamically amplified. As a consequence, increasing the
memory horizon does not merely improve a numerical approximation in a passive
way; it may substantially extend the time interval over which the truncated
trajectory remains coherent with the full-history discrete GL reference. This motivated the
definition of the coherence-loss time \(T_{\rm loss}\) and of the critical
memory horizon \(T_m^{\rm crit}\).

The second main finding is that the memory horizon required to preserve
coherence depends strongly on the dynamical regime. In the \((f,T_m)\) plane,
the critical memory threshold varies non-monotonically with the forcing
amplitude, with extended regions where much longer memory horizons are needed
to preserve coherence. A complementary scan in the \((q,T_m)\) plane showed
that the fractional order also plays a decisive role. The dependence on \(q\)
is not simply monotonic: instead, certain ranges of fractional order amplify
the effect of memory truncation more strongly than others. These results show
that the admissible truncation length cannot be selected only from the
algebraic decay of the GL coefficients, but must also take into account the
sensitivity of the nonlinear dynamics.

The third main finding concerns the local spectral structure of the truncated
GL kernel. We derived the characteristic equation associated with the
finite-memory GL operator and showed that the truncated kernel admits a local
delay-type interpretation. In particular, a low-order matching yields an
effective representation in terms of an instantaneous contribution plus a
delayed exponential term, providing a causal local surrogate of the truncated
memory kernel. This representation should not be interpreted as a global
reduction of the finite-memory fractional system to a single delayed equation.
Rather, it serves as a local spectral approximation that helps clarify the
connection between truncated fractional memory and delay-type dynamics.

Finally, we quantified the effective delay complexity of the truncated GL
kernel by approximating it with a finite sum of positive-delay exponentials.
The resulting map of \(r_{\min}(q,T_m)\) shows that, although the truncated
kernel admits a simple local delay-type surrogate, its spectral structure is
generally richer than that of a single delayed contribution. The number of
positive-delay modes required to reproduce its local spectral action depends
jointly on the fractional order and on the retained memory horizon. This
provides an operational measure of the delay-like complexity, or
compressibility, of the truncated memory operator.

From a practical point of view, the results are relevant for numerical
simulations of fractional nonlinear systems in which memory truncation is used
to reduce computational cost. The analysis shows that choosing a memory
horizon involves a trade-off between efficiency and dynamical reliability. A
short memory horizon reduces storage and computational effort, but it may
discard a part of the memory kernel that is dynamically relevant. The
coherence-loss time and the critical memory horizon provide practical
diagnostics for selecting a truncation length compatible with a prescribed
tolerance and observation time.

The present study also has limitations. The thresholds \(T_m^{\rm crit}\), the
coherence-loss time \(T_{\rm loss}\), and the delay-complexity measure
\(r_{\min}\) are operational quantities. They depend on the prescribed
tolerances, the observation window, the time step, the chosen spectral domain,
and the parameter ranges explored. Similarly, the local spectral residual used
to identify transition regions should not be interpreted as an exact global
bifurcation condition for the nonlinear periodically forced system. The
positive-delay representation is a local spectral approximation of the
truncated GL kernel, not a proof of global equivalence between the fractional
Duffing oscillator and a finite-delay differential equation.

Several directions for future work follow naturally from these results. First,
the dependence of \(T_m^{\rm crit}\) on the tolerance, the integration step,
and the observation time should be studied systematically. Second, the
heuristic amplification picture based on the discarded memory tail could be
developed into a more quantitative theory by estimating the effective
sensitivity rate of the nonlinear fractional dynamics. Third, the present
framework could be extended to other fractional orders, other nonlinear
oscillators, coupled systems, and fractional chaotic models. Finally, the
positive-delay representation could be compared with other kernel-compression
or rational-approximation techniques in order to better understand when a
finite-memory fractional kernel can be efficiently represented by a small
number of causal modes.

Overall, the results show that finite-memory approximations of fractional
damping should be interpreted dynamically, not only numerically. The retained memory horizon affects both the accuracy of the kernel representation and the coherence time of the resulting nonlinear dynamics. The positive-delay
representation provides a useful bridge between fractional memory and
delay-type dynamics, but its validity is local and spectral rather than a
global equivalence with a finite-delay system.

{
Another natural extension of the present work is to compare the GL-based
finite-memory analysis developed here with formulations arising in discrete
fractional calculus, where memory effects are incorporated directly at the
discrete level and additional structural properties may become available.
}

\section*{Data and code availability}

The MATLAB codes and data supporting the results of this study will be made
available in a public repository upon publication.

\section{Acknowledgments}

This work was supported by the Spanish State Research Agency (AEI) and the European Regional Development Fund (ERDF,EU) under Project No. PID2023-148160NB-I00 (MCIN/AEI/10.13039/501100011033).

\end{document}